\documentclass[twocolumn]{aastex631}

\usepackage{amsmath}
\usepackage{graphicx}
\usepackage{longtable}
\usepackage{academicons}
\definecolor{orcidlogocol}{HTML}{A6CE39}

\definecolor{myblue}{RGB}{0, 160, 240} 
\definecolor{mygreen}{RGB}{0, 180, 0}

\defcitealias{Breuval2021}{B21}
\defcitealias{Macri2015}{M15}
\defcitealias{Anderson2013}{A13}
\defcitealias{Kovtyukh2008}{K08}
\defcitealias{Laney2007}{LC07}
\defcitealias{Sziladi2007}{S07}
\defcitealias{Acharova2012}{A12}
\defcitealias{Fernie1995}{F95}
\defcitealias{Owens2022}{OW22}

\received{11 May 2022}
\revised{15 September 2022}
\accepted{9 November 2022}

\shorttitle{An improved calibration of the wavelength dependence of the metallicity effect}
\shortauthors{Breuval L.}

\begin{document}

\title{An Improved Calibration of the Wavelength Dependence of Metallicity \\ on the Cepheid Leavitt law}

\author[0000-0003-3889-7709]{Louise Breuval }
\affiliation{Department of Physics and Astronomy, Johns Hopkins University, Baltimore, MD 21218, USA}
\affiliation{LESIA, Observatoire de Paris, Universit\'e PSL, CNRS, Sorbonne Universit\'e, Universit\'e Paris Cit\'e, \\ 5 place Jules Janssen, 92195 Meudon, France}
\email{lbreuva1@jhu.edu}

\author[0000-0002-6124-1196]{Adam G. Riess}
\affiliation{Department of Physics and Astronomy, Johns Hopkins University, Baltimore, MD 21218, USA}
\affiliation{Space Telescope Science Institute, 3700 San Martin Drive, Baltimore, MD 21218, USA}

\author[0000-0003-0626-1749]{Pierre Kervella}
\affiliation{LESIA, Observatoire de Paris, Universit\'e PSL, CNRS, Sorbonne Universit\'e, Universit\'e Paris Cit\'e, \\ 5 place Jules Janssen, 92195 Meudon, France}

\author[0000-0001-8089-4419]{Richard I. Anderson}
\affiliation{Institute of Physics, Laboratory of Astrophysics, \'Ecole Polytechnique F\'ed\'erale de Lausanne (EPFL), \\ Observatoire de Sauverny, 1290 Versoix, Switzerland}

\author[0000-0002-5527-6317]{Martino Romaniello}
\affiliation{European Southern Observatory, Karl-Schwarzschild-Strasse 2, 85478 Garching bei München, Germany}


\begin{abstract}

The Cepheid period-luminosity (PL) relation (or Leavitt law) has served as the first rung of the most widely used extragalactic distance ladder and is central to the determination of the local value of the Hubble constant ($H_0$). We investigate the influence of metallicity on Cepheid brightness, a term that significantly improves the overall fit of the distance ladder, to better define its wavelength dependence. To this aim, we compare the PL relations obtained for three Cepheid samples having distinct chemical composition (in the Milky Way and Magellanic Clouds) and focusing on the use of improved and recent data while covering a metallicity range of about 1 dex. We estimate the metallicity effect (hereafter $\gamma$) in 15 filters from mid-IR to optical wavelengths, including five Wesenheit indices, and we derive a significant metallicity term in all filters, in agreement with recent empirical studies and models, in the sense of metal-rich Cepheids being brighter than metal-poor ones. We describe the contribution of various systematic effects in the determination of the $\gamma$ term. We find no evidence of $\gamma$ changing over the wavelength range $0.5-4.5 \, \rm \mu m$, indicating that the main influence of metallicity on Cepheids is in their luminosity rather than color. Finally, we identify factors that sharpen the empirical constraints on the metallicity term over past studies, including corrections for the depth of the Magellanic Clouds, better-calibrated Cepheid photometry, improved Milky Way extinction estimates, and revised and expanded metallicity measurements in the LMC. \\
~ \\

\end{abstract}


\section{Introduction} 
\label{sec:intro}

The Cepheid period-luminosity (PL) relation \citep[or Leavitt law,][]{Leavitt1912} is a fundamental tool for measuring astronomical distances and has been used for decades to estimate the current expansion rate of the Universe, the Hubble constant \citep{Freedman2001, Riess2022}. Recently, a significant tension of $5\sigma$ has arisen between the prediction of $H_0$ from the Cosmic Microwave Background (CMB) data from \citet{Planck2018} assuming a $\Lambda$CDM model, $H_0 = 67.4 \pm 0.5 \, \rm km \, s^{-1} \, Mpc^{-1}$, and its empirical estimate based on Cepheids and SNe Ia measurements, $H_0 = 73.04 \pm 1.04 \, \rm km \, s^{-1} \, Mpc^{-1}$ \citep{Riess2022}. The persistence of this discrepancy could have significant implications in cosmology as it may suggest a breach in the standard model \citep{DiValentino2021}. It is therefore important to empirically scrutinize the nature of the Leavitt Law.

In the present paper, we aim at measuring the influence of chemical abundance on Cepheid brightness, as this term has been found to significantly improve the quality of the fit of the distance ladder.
The difference in metallicity between Cepheids used to calibrate the PL relation and Cepheids in SNe Ia host galaxies is usually taken into account by including a corrective term ($\gamma$) in the PL relation, such that: 
\begin{equation}
\label{eq:PLZ}
M = \alpha \, (\log P - \log P_0) + \delta + \gamma \, \rm [Fe/H]
\end{equation}
The majority of extragalactic Cepheids in SN Ia hosts have inferred abundances that are similar to those in two distance ladder anchors, the Milky Way (MW) and NGC 4258, in terms of metal content \citep[see Fig. 21 in][]{Riess2022}. However, the Large Magellanic Cloud (LMC) which contains more metal-poor Cepheids, is also often used as an anchor in the distance ladder \citep{Riess2019}: its distance was measured with high precision by \citet{Pietrzynski2019} using detached eclipsing binaries (DEBs). The improved characterization of the distance ladder depends on an improved constraint on the metallicity term to span such a range of metallicity.  The best accuracy on this term can thus be obtained by using an even larger metallicity range including the even more metal poor Small Magellanic Cloud (SMC) and its recently measured DEB distance  \citep{Graczyk2020}.

Various estimates of the metallicity effect were published in the last two decades, based on different methods, samples, photometry, distances or chemical abundances. They are listed in Table \ref{table:gamma_lit}. Early studies using nonlinear convecting models \citep{Bono1999, Bono2008, Caputo2000, Marconi2005} based on masses and luminosities provided by stellar evolutionary calculations predicted a positive metallicity effect ($\gamma > 0$), meaning that metal-rich Cepheids are fainter than metal-poor ones. A positive metallicity term might be explained as a strictly atmospheric effect from line blanketing with higher metallicity producing more absorption lines to decrease the flux emitted by the star and make it appear fainter than expected in the optical \citep{Freedman2011}, although this explanation does not address possible changes to a Cepheid's bolometric luminosity. More recently, \citet{Anderson2016} used stellar models from the Geneva group including the effects of rotation and derived a strong negative dependence in the optical and NIR (see Sect. \ref{sec:comp_lit}), meaning that metal-rich Cepheids would be brighter \citep[see also][]{DeSomma2022}. 

On the other hand, almost all recent empirical studies have obtained a negative metallicity term ($\gamma < 0$) based on extragalatic Cepheids \citep{Macri2006, Scowcroft2009}, Baade-Wesselink distances of Milky Way and Magellanic Cloud Cepheids \citep{Storm2011b, Groenewegen2013, Gieren2018}, DEB distances for Magellanic Cloud Cepheids \citep{Wielgorski2017}, \textit{Gaia} \citep{GaiaMission2016} DR2 or EDR3 parallaxes \citep{Groenewegen2018, Ripepi2020, Ripepi2021, Ripepi2022, Riess2021}. \citet{Romaniello2008} provided a range of metal abundances of individual LMC Cepheids which were used by \citet{Freedman2011} to derive a negative metallicity term in the mid-infrared (MIR) \textit{Spitzer} bands, becoming progressively weaker and then positive towards optical wavelengths, with a crossover around the near infrared (NIR). However, a significant revision as well as an expansion of these measurements by \citet{Romaniello2021} concluded there was no significant differences among the individual metallicities of LMC Cepheids, negating the ability to constrain the metallicity term internal to the LMC. Non-trivial depth of the Magellanic Clouds is also an important factor as the scale of the metallicity term between the Clouds and the Milky Way is $<$ 0.1 mag.



Improving data quality necessitates a study of the metallicity term outside the context of the distance ladder and over a broader range of wavelength. Because the accuracy required to resolve the metallicity term, a few hundredths of a magnitude, is comparable to historic zeropoint errors, consistent calibration of Cepheid photometry is paramount.
While the distance ladder combines constraints on the metallicity term from both metallicity gradients within hosts and abundance differences between hosts \citep{Riess2022}, we focus here on the constraints from the latter as these offer the best combination of simplicity, wavelength coverage and constraining power.  

In \citet{Breuval2021} (hereafter \citetalias{Breuval2021}), we calibrated $\gamma$ in 7 ground-based filters covering NIR and optical wavelengths, including two Wesenheit indices. The present work aims at improving, expanding and complementing this preliminary study with new data, in particular by including two additional mid-infrared \textit{Spitzer} bands, three additional optical \textit{Gaia} bands as well as three supplementary Wesenheit indices \citep[including the HST reddening-free $W_H$ band used in the SH0ES papers, e.g.][]{Riess2022}, resulting in a total of 15 different filters. The large wavelength coverage ($0.5 < \lambda < 4.5 \, \rm \mu m$) also allows for a study of a possible dependence between $\gamma$ and $\lambda$. Following \citetalias{Breuval2021}, we adopt three Cepheid samples of different chemical abundance which also have precise distances: Milky Way, LMC and SMC Cepheids. Sect. \ref{sec:data} describes the data used in this analysis. The method is outlined in Sect. \ref{sec:method} and the results are given in Sect. \ref{sec:results}. We discuss our findings in Sect. \ref{sec:discussion} and conclude with perspectives in Sect. \ref{sec:conclusion}. \\

\section{Data} 
\label{sec:data}

This section describes the catalogs used in this analysis (photometry, reddenings, distances and metallicities), they are listed in Table \ref{table:photom_ZP}.  \\

\begin{longtable*}{c c c c c c c c}
\caption{Empirical and theoretical estimates of the metallicity effect ($\gamma$ in mag/dex) on Cepheid brightness. }\\
\hline
\hline
Band & $\gamma$ & Reference & Method &  \\	
\hline
\endfirsthead
\multicolumn{5}{c}{\textbf{Table \ref{table:gamma_lit}} (continued)} \\
\hline
Band & $\gamma$ & Reference & Method &  \\	
\hline 
\endhead
\hline
\endfoot
\endlastfoot
\multicolumn{5}{c}{\bf{THEORETICAL STUDIES}}  \\
\hline
$V$ & $+0.40$ & \citet{Bono1999} & Nonlinear convecting models & [Fe/H] \\
$K$ & $-0.08$ & &  &   \\
\hline
$V$, $I$ & $+0.27$ & \citet{Caputo2000} & Nonlinear convecting models  & [Fe/H] \\
\hline
$V$, $I$ & $\gamma > 0$  & \citet{Marconi2005} & Nonlinear convecting models  & [Fe/H]  \\
\hline
$W_{VI}$ & $+0.05 \pm 0.03$ & \citet{Bono2008} & Nonlinear convecting models  & [Fe/H]  \\
\hline
$V$	& $-0.277 \pm 0.102$ & \citet{Anderson2016} & Geneva evolution models including &   \\
$H$	& $-0.214 \pm 0.086$ &  &  the effects of rotation ($2^{\rm nd}$ crossing),  &   \\
$W_{VI}$& $-0.221 \pm 0.097$ &  & average between blue and red edge &   \\
$W_{H}$	& $-0.205 \pm 0.084$ &  &  &   \\
\hline  
$W_G$ & $-0.13$ to $-0.25$ & \citet{DeSomma2022} & Nonlinear convecting models  &   \\
$W_H$ & $-0.13$ to $-0.19$ &  &   &   \\
$W_{VI}$ & $-0.15$ to $-0.17$ &  &   &   \\
$W_{VK}$ & $-0.14$ to $-0.18$ &  &   &   \\
\hline

\multicolumn{5}{c}{\bf{EMPIRICAL STUDIES}}  \\
\hline
$W_{VI}$ & $-0.24 \pm 0.16$ & \citet{Kennicutt1998} & 2 fields in M101 & [O/H]  \\
\hline
$W_{VI}$ & $-0.24 \pm 0.05$ & \citet{Sakai2004} & TRGB/Cepheid distances to nearby galaxies  & [O/H]  \\
\hline
$W_{VI}$ & $-0.29 \pm 0.10$ & \citet{Macri2006} & 2 fields in NGC 4258  & [O/H]  \\ 
\hline
$K$ & $\sim 0$ & \citet{Romaniello2008} & MW, LMC, SMC + HR spectra  & [Fe/H]  \\
$V$ & $\gamma > 0$ &   &   \\
\hline
$W_H$ & $-0.23 \pm 0.17$ & \citet{Riess2009} & HST photometry, NGC 4258 & [O/H]  \\
\hline
$W_{VI}$ & $-0.29 \pm 0.11$ & \citet{Scowcroft2009} & 4 fields in M33  & [O/H] \\ 
\hline
$W_H$ & $-0.10 \pm 0.09$ & \citet{Riess2011} & HST photometry, MW, LMC, NGC 4258 & [O/H]  \\
\hline
$V$ &$+0.50 \pm 0.31$ & \citet{Freedman2011}  &  Abundances of individual LMC Cepheids & [Fe/H]  \\
$J$ & $+0.14 \pm 0.07$ & &  & \\
$H$ & $+0.05 \pm 0.02$ & & &  \\
$K$ & $+0.02 \pm 0.03$ & \\
$[3.6\, \rm \mu m]$ & $-0.39 \pm 0.16$ & & & \\
$[4.5\, \rm \mu m]$ & $-0.25 \pm 0.18$ & & & \\
$[5.8\, \rm \mu m]$ & $-0.39 \pm 0.17$ & & & \\
$[8.0\, \rm \mu m]$ & $-0.38 \pm 0.16$ & &  & \\ 
\hline
$[3.6\, \rm \mu m]$ & $-0.09 \pm 0.29$ & \citet{Freedman2011b} & MW, LMC, SMC & [Fe/H] \\
\hline
$V$ & $+0.09 \pm 0.10$ &  \citet{Storm2011b} & MW, LMC, SMC + IRSB BW distances  & [Fe/H] \\
$I$ & $-0.06 \pm 0.10$ & &  $p = 1.55 - 0.186 \log P$  \\
$W_{VI}$ & $-0.23 \pm 0.10$ & & \citep{Storm2011a}   \\
$J$ & $-0.10 \pm 0.10$ & & \\
$K$ & $-0.11 \pm 0.10$ & &  \\
$W_{JK}$ & $-0.10 \pm 0.10$ & & \\
\hline
$V$ & $+0.23 \pm 0.11$ & \citet{Groenewegen2013} & MW, LMC, SMC + IRSB BW distances  & [Fe/H] \\
$K$ & $-0.05 \pm 0.10$ & &   $p = 1.50 - 0.24 \log P$  \\
$W_{VK}$ & $+0.04 \pm 0.10$ &  & \\ 
\hline
$W_H$ & $-0.14 \pm 0.06$ & \citet{Riess2016} & HST photometry, MW, LMC, NGC 4258 & [O/H]  \\
\hline
$V$ & $-0.022 \pm 0.076$ & \citet{Wielgorski2017} & LMC, SMC + DEB distances   & [Fe/H] \\
$I$ & $-0.015 \pm 0.071$ & \\
$J$ & $-0.042 \pm 0.069$ & \\
$H$ & $-0.012 \pm 0.069$ & \\
$K$ & $-0.017 \pm 0.069$ &  \\
$W_{VI}$ & $-0.025 \pm 0.067$ &  \\
$W_{JK}$ & $-0.022 \pm 0.067$ & \\
\hline
$V$ & $-0.238 \pm 0.186$ & \citet{Gieren2018} & MW, LMC, SMC + IRSB BW distances  & [Fe/H] \\
$I$ &  $-0.293 \pm 0.150$ & &  $p = 1.55 - 0.186 \log P$ \citep{Storm2011a}  \\
$W_{VI}$ & $-0.335 \pm 0.059$ & &    \\
$J$ & $-0.270 \pm 0.108$ & &   \\
$K$ & $-0.232 \pm 0.064$ & &  \\
$W_{JK}$ & $-0.221 \pm 0.053$ &  \\
\hline
$V$ & $-0.041 \pm 0.260$ & \citet{Groenewegen2018} & MW \textit{Gaia} DR2 parallaxes, ZP $= -0.046\, \rm mas$ & [Fe/H]  \\
$K$ & $-0.168 \pm 0.146$ &  \\
$W_{VK}$ & $-0.188 \pm 0.142$ &  \\
\hline
$W_H$ & $-0.17 \pm 0.06$ & \citet{Riess2019} & HST photometry, MW, LMC, NGC 4258 & [O/H]  \\
\hline
$K$ & $-0.082 \pm 0.138$ & \citet{Ripepi2020} & MW \textit{Gaia} DR2 parallaxes, ZP $= -0.049\, \rm mas$  & [Fe/H] \\
$W_{JK}$ & $-0.284 \pm 0.115$ &  \\
\hline
$K$ & $-0.456 \pm 0.099$ & \citet{Ripepi2021} & MW \textit{Gaia} EDR3 parallaxes + HR spectra  & [Fe/H] \\
$W_{JK}$ & $-0.465 \pm 0.071$ &  \\
$W_{VK}$ & $-0.459 \pm 0.107$ &  \\
\hline
$V$ & $-0.048 \pm 0.055$ & \citet{Breuval2021} & MW, LMC, SMC & [Fe/H]  \\  
$I$ & $-0.138 \pm 0.053$ &  & \textit{Gaia} EDR3 parallaxes + DEB distances    \\
$W_{VI}$ & $-0.251 \pm 0.057$ &    \\
$J$ & $-0.208 \pm 0.052$ &    \\
$H$ & $-0.152 \pm 0.092$ &    \\
$K$ & $-0.221 \pm 0.051$ &    \\
$W_{JK}$ & $-0.214 \pm 0.057$ &    \\
\hline
$W_H$ & $-0.217 \pm 0.046$ & \citet{Riess2022} & Cepheids with HST/WFC3 photometry & [O/H] \\ 
\hline
$W_G$ & $-0.520 \pm 0.090$ & \citet{Ripepi2022} & MW \textit{Gaia} EDR3 parallaxes + HR spectra & [Fe/H] \\
\hline
~ \\
\label{table:gamma_lit}
\end{longtable*} 

\newpage

\subsection{Photometry}
\label{sec:photometry}

In order to minimize systematic uncertainties related to the use of disparate photometric systems present in many literature compilations of Cepheid magnitudes, we adopt the most homogeneous and consistently calibrated data sets available, while ensuring at the same time that the best light curve coverage is obtained for the Cepheids.  \\

\textbf{NIR ground $J$, $H$ and $K$ filters}: in the NIR, we adopted the photometry from \citet{Monson2011} transformed in the 2MASS system for MW Cepheids. Mean magnitudes of LMC Cepheids are taken from the VISTA survey for the Magellanic Clouds (VMC) by \citet{Ripepi2022VMC} in $J$ and $K$ and from the Synoptic Survey by \citet{Macri2015} (hereafter \citetalias{Macri2015}) in $H$, the latter also includes additional Cepheids from \citet{Persson2004}. Finally, the VMC survey by \citet{Ripepi2017} provides $J$ and $K$ band light curves for a large number of SMC Cepheids. We complemented these data with the $H$ band single epoch photometry from the \citet{Kato2007} Point Source Survey. In the LMC and SMC, VMC mean magnitudes are converted into the 2MASS system using the transformations from \citet{Gonzalez2018}. Finally, empirically-derived transformations for zeropoints and color terms are applied to LMC and SMC NIR photometry to match the 2MASS system, the details can be found in Sect. 2.2 and 2.3 in \citetalias{Breuval2021}. \\

\textbf{Optical ground $V$ and $I$ filters}: in the $V$ and $I$ bands, we adopt the compilation of light curves from \citet{Berdnikov2008} for MW Cepheids. In the LMC and SMC, we adopt the OGLE-IV survey from \citet{Soszynski2015}, combined with the Shallow Survey of bright LMC Cepheids by \citet{Ulaczyk2013}. For consistency, we select the list of Cepheids from \citet{Macri2015} for the LMC sample.   We note the lack of a more consistently calibrated, modern set of all-sky optical Cepheid data for the Milky Way sample, making it the most limiting dataset in the following studies.  \\

\textbf{\textit{Gaia} optical $G$, $BP$ and $RP$ filters}: We used the intensity averaged mean magnitudes provided in the \textit{Gaia} DR3 \texttt{"vari\_cepheid"} catalog by \citet{Ripepi2022Gaia} in the three optical \textit{Gaia} bands, for Milky Way, LMC and SMC Cepheids. For the LMC and SMC samples we adopted all Cepheids in the regions defined in Table 1 by \citet{Ripepi2022Gaia}. As for Milky Way Cepheids we considered that ”AllSky” Cepheids are those outside of the regions of the LMC, SMC, M31 and M33 \citep{Ripepi2022Gaia}. These mean magnitudes are internally photometrically consistent. All stars have at least 15 epochs in \textit{Gaia} bands and an average of 45 epochs.  \\

\textbf{\textit{Spitzer} mid-infrared [3.6 $\, \rm \mu$m] and [4.5 $\, \rm \mu$m] filters}: A sample of 37 Galactic Cepheids were observed with the \textit{Spitzer} Space Telescope, their mean magnitudes are provided in the mid-infrared $[3.6 \, \rm \mu m]$ and $[4.5 \, \rm \mu m]$ filters by \citet{Monson2012}. Similarly, 85 LMC Cepheid and 90 SMC Cepheid mean magnitudes were measured in the same filters by \citet{Scowcroft2011} and \citet{Scowcroft2016} respectively. These are internally photometrically consistent. \\

\begin{table*}[t!]
\footnotesize
\caption{References for the data adopted in this study.}
\centering
\begin{tabular}{ c |  c  c c}
\hline
\hline
 & Milky Way & Large Magellanic Cloud & Small Magellanic Cloud \\
\hline
$V, I$ & \citet{Berdnikov2008} & \citet{Soszynski2015} & \citet{Soszynski2015}  \\
 &  & \citet{Ulaczyk2013} &   \\
\hline
$J, H, K$ & \citet{Monson2011}  & $J, \, K$: \citet{Ripepi2022VMC} & $J, \, K$: \citet{Ripepi2017}   \\
   &  & $H$:  \citet{Macri2015}, & $H$: \citet{Kato2007} \\
   &  &  \citet{Persson2004} \\
\hline
$G, BP, RP$ & \textit{Gaia} DR3  & \textit{Gaia} DR3 & \textit{Gaia} DR3  \\
   & \citep{Ripepi2022Gaia} & \citep{Ripepi2022Gaia} & \citep{Ripepi2022Gaia} \\
\hline
$[3.6 \, \rm \mu m]$, $[4.5 \, \rm \mu m]$  & \citet{Monson2012} & \citet{Scowcroft2011} & \citet{Scowcroft2016}   \\
\hline
$F160W, \, F555W, F814W$  & \citet{Riess2021} & \citet{Riess2019} & $-$  \\
\hline
Reddening & (a) Bayestar dust map \citep{Green2019}  & \citet{Skowron2021} & \citet{Skowron2021}  \\
 	& (b) Period-color relation \citep{Riess2022}  & reddening maps & reddening maps \\
 	& (c) SPIPS method \citep{Trahin2021}  & &  \\
\hline
Distance &  \citet{BailerJones2021} distances & DEB distance  & DEB distance  \\
 & (includes ZP correction by		& $49.59 \pm 0.09 \pm 0.54 \, \rm kpc$ & $62.44 \pm 0.47 \pm 0.81\, \rm kpc$  \\
 & \citet{Lindegren2021_plx_bias}) & \citep{Pietrzynski2019} &\citep{Graczyk2020}   \\
 & + additional ZP of $0.014 \, \rm mas$ & + geometry correction & + geometry correction \\
 &   \citep{Riess2021} &  &  \\
\hline
Metallicity   & \citet{Genovali2014, Genovali2015} & \citet{Romaniello2021} & \citet{Gieren2018} \\
  & $\rm [Fe/H] = +0.088 \, \rm dex \, \, (\sigma = 0.022)$ & $\rm [Fe/H] = -0.407 \pm 0.020 \, dex$ & $\rm [Fe/H] = -0.75 \pm 0.05 \, dex$ \\
  & (depends on the sample)  &  &  \\
\hline
\multicolumn{4}{c}{ }  \\
\multicolumn{4}{c}{ }  \\
\end{tabular}
\label{table:photom_ZP}
\end{table*}

\textbf{\textit{Hubble} Space Telescope (HST) Wide Field Camera 3 (WFC3) filters}: The HST WFC3 filters $F555W$, $F814W$ and $F160W$ are particularly interesting since they can be combined into the reddening-free Wesenheit index $W_H$, which is also used to observe extragalactic Cepheids \citep{Riess2022}, cancelling photometric zero-point errors on the distance scale. \citet{Riess2021} provides HST/WFC3 photometry for 75 Milky Way Cepheids obtained by the spatial scanning technique. Similarly, \citet{Riess2019} measured HST/WFC3 mean magnitudes in the same filters for 70 LMC Cepheids. Unfortunately, there is currently no available HST photometry for SMC Cepheids: in the HST photometric system, the analysis will be limited to Galactic and LMC Cepheids only. \\

\textbf{Systematic uncertainties}: In $V$, $I$, $J$, $H$ and $K$, the photometry for Galactic and Magellanic Cloud Cepheids was obtained with different instruments and sometimes in different systems (see Table \ref{table:photom_ZP}): we include an error of $0.020 \, \rm mag$ to the PL intercepts in these five filters for each host. For \textit{Gaia}, \textit{Spitzer} and \textit{HST} photometry, the data are taken with the same instruments and reduced by the same teams for the 3 galaxies so they do not require any systematic zeropoint uncertainty.

Additionally, the \citet{Berdnikov2008} catalog gathers observations made between 1986 and 2004 which are rather inhomogeneous, therefore we include quadratically an additional photometric zero-point uncertainty of 0.010 mag for MW Cepheids in $V$ and $I$, which sums to a total systematic zeropoint difference between Cepheids in the MW and either Cloud of $\sigma=0.03$ mag in these bands. For the LMC sample, \citetalias{Macri2015} reports zero-point differences of $0.018 \pm 0.067$, $-0.016 \pm 0.058$ and $0.000 \pm 0.054$ mag in $J$, $H$, $K$ respectively between \citetalias{Macri2015} and \citet{Persson2004}, after transformation into the 2MASS system. We adopt these values as photometric zero-point error in the NIR for LMC Cepheids to account for the internal consistency of the \citetalias{Macri2015} catalog. Finally, from the comparison between \citet{Kato2007} and VMC magnitudes in the SMC, a photometric zero-point uncertainty of $0.010 \, \rm mag$ is adopted in the $J$, $H$, $K$ bands for the Cepheids of this galaxy. \\

\subsection{Reddening}

Apparent magnitudes must be carefully corrected for extinction, due to the presence of dust on the line of sight, using a reddening law and consistent $E(B-V)$ values.  Past studies have relied on the \citet{Fernie1995} database which is an inhomogeneous compilation of major colour excess determinations published since 1975 derived from 17 sources of mostly photoelectric data in non-standard bandpass systems and which are therefore inadequate for providing consistent reddening estimates with an accuracy of a few hundreths of a magnitude across the sky needed for this study.

We make use of three different sources of reddening values for Milky Way Cepheids. The first one is the 3D dust maps by \citet{Green2019} based on homogeneous \textit{Gaia}, Pan-STARRS 1 and 2MASS photometry. As a second method, we derive reddening values by comparing the observed color $(V-I)_{\rm obs}$ of Cepheids with their intrinsic color $(V-I)_{\rm intr}$ obtained from the period-color relation: $(V-I)_{\rm intr} = (0.25 \pm 0.01) \log P + (0.50 \pm 0.01)$ \citep{Riess2022}. Finally, we adopt as third estimate the reddening values from \citet{Trahin2021} obtained with the SPIPS algorithm \citep{Merand2015} for MW Cepheids having an optimal set of spectro-, photo- and interferometric data. In the Monte Carlo sampling procedure described in Sect. \ref{sec:PLZrelation}, $E(B-V)$ values are selected randomly among these three catalogs for each star and the procedure is repeated over 10,000 iterations, which allows us to account for the covariance of these methods. Finally, in the LMC and SMC we used the reddening maps by \citet{Skowron2021} and we transform $E(V-I)$ into $E(B-V)$ using the relation adopted by \citet{Skowron2021}: $E(V-I) = 1.237 \times E(B-V)$. Milky Way Cepheids are particularly affected by interstellar reddening: the stars of our MW sample have a mean $E(B-V)$ of 0.5 mag (dispersion, $\sigma = 0.3$ mag), while in the LMC they have a mean $E(B-V)$ of 0.11 mag ($\sigma = 0.05$ mag) and of 0.05 mag ($\sigma = 0.02$ mag) in the SMC. Reddening uncertainties are propagated through uncertainty in the reddening law in the next section.\\

\begin{table*}[t!]
\caption{Filters in which the effect of metallicity is calibrated in this study: effective central wavelength ($\lambda^{\rm eff}_0$) from the SVO filter profile service, ratios of total-to-selective absorption ($R_{\lambda}$) from \citet{Fitzpatrick1999} assuming $R_V = 3.1 \pm 0.1$ (and from \citet{Indebetouw2005} for \textit{Spitzer} bands), width of the instability strip (WIS) and mean metallicity of the MW sample.}
\centering
\begin{tabular}{c c c c c}
~\\
\hline
\hline
Filter & $\lambda^{\rm eff}_0$ & $R_{\lambda}$ & WIS & [Fe/H]$_{\rm MW}$  \\
 & ($\rm \mu m$) & & (mag)  & (dex) \\
\hline
$BP$ 	& 0.5036	& $\,\,3.433 \pm 0.111\,\,$ & 0.23   & $0.087 \pm 0.056$    \\
$V$ 	& 0.5468	& $\,\,3.057 \pm 0.099\,\,$ & 0.22   & $0.093 \pm 0.058$  \\
$G$ 	& 0.5822	& $\,\,2.783 \pm 0.090\,\,$ & 0.19   & $0.087 \pm 0.056$    \\
$RP$ 	& 0.7620	& $\,\,1.831 \pm 0.059\,\,$ & 0.16   & $0.087 \pm 0.056$  \\
$I_C$  	& 0.7829	& $\,\,1.777 \pm 0.057\,\,$ & 0.14   & $0.097 \pm 0.064$   \\
$J$ 	& 1.2350	& $\,\,0.812 \pm 0.026\,\,$ & 0.11   & $0.094 \pm 0.081$  \\
$H$ 	& 1.6620	& $\,\,0.508 \pm 0.016\,\,$ & 0.09   & $0.094 \pm 0.082$   \\
$K$	 	& 2.1590	& $\,\,0.349 \pm 0.011\,\,$ & 0.07   & $0.093 \pm 0.087$   \\
$[3.6 \, \rm \mu m]$ & 3.5075	& $\,\,0.198 \pm 0.023\,\,$ & 0.07  & $0.146 \pm 0.075$   \\
$[4.5 \, \rm \mu m]$ & 4.4366	& $\,\,0.152 \pm 0.028\,\,$ & 0.07  & $0.146 \pm 0.075$  \\ 
\hline
\multicolumn{5}{c}{Wesenheit indices}  \\
\hline
$W_{G} $ & \multicolumn{2}{c}{$ G - 1.900 \, (BP - RP)$} & 0.10                 & $0.087 \pm 0.056$  \\
$W_{VI}$ &  \multicolumn{2}{c}{$ I - 1.387 \,(V-I)$} & 0.077             & $0.098 \pm 0.065$  \\
$W_{JK}$ & \multicolumn{2}{c}{$ K - 0.735 \,(J-K)$} & 0.086              & $0.094 \pm 0.088$  \\
$W_{VK}$ & \multicolumn{2}{c}{$ K - 0.127 \,(V-K)$} & 0.077              & $0.098 \pm 0.090$  \\
$W_{H}$  & \multicolumn{2}{c}{$ F160W- 0.386 \,(F555W-F814W)$} & 0.069   & $0.099 \pm 0.089$  \\
\hline
~ \\
\end{tabular}
\label{table:filters}
\end{table*}

\subsection{Metallicity}

For Milky Way Cepheids, we adopted in priority the iron abundances by \citet{Genovali2015} and complemented these values with the catalog by \citet{Genovali2014}. The latter also provides additional abundances from the literature for 375 other Galactic Cepheids, for which we set the uncertainty to $0.1 \, \rm dex$. All metallicity measurements provided in these catalogs are rescaled to the same solar abundance. The average metallicity of our full MW sample is $+0.088 \, \rm dex$ with a dispersion of $0.122 \, \rm dex$. However, the mean metallicity of MW Cepheids can differ depending on the sample (e.g. Cepheids for which we have optical photometry, vs those for which we have NIR photometry), therefore we consider the mean metallicity of the exact sample used in each filter. These mean values are similar in all filters (from $+0.087 \, \rm dex$ to $+0.099 \, \rm dex$), except in the two Spitzer bands where the mean metallicity is slightly higher ($+0.146 \, \rm dex$), which can be explained by the small size of the sample. In order to derive the uncertainties for the mean MW metallicity in each filter, we run a bootstrap algorithm on the available metallicity values and we adopt the 99\% confidence interval ($3\sigma$) of the distribution of the mean values.  Considering the limited size of our metallicity sample, particularly in the Spitzer bands, the bootstrapping approach enables us to determine the confidence interval of the mean metallicity without assumption on the normality of the metallicity sample distribution \citep{Efron1986}. These values are listed in Table \ref{table:filters}.

We note that additional constraining power is available by retaining the individual MW abundances in the analysis \citep{Riess2021} but we have chosen the simpler host-to-host analysis for its transparency, although it relies on the use of a single average MW metallicity.


\citet{Romaniello2021} recently obtained high resolution spectra for 89 Cepheids in the LMC and derived their chemical abundances and revised the measurements of those from \citet{Romaniello2008}. They concluded that they are consistent within the errors with a single common abundance of $-0.409\, \rm dex$ with a dispersion of $0.076 \, \rm dex$ (similar to the uncertainty per measurement of 0.07 dex), which is more metal-poor by 0.1 dex and the breadth of the distribution is half as wide (see discussion in \citet{Romaniello2021}. We adopt this mean value for all LMC Cepheids with an uncertainty of $0.02 \, \rm dex$. In the SMC, we follow \citet{Gieren2018} and adopt a mean metal abundance of $-0.75 \pm 0.05 \, \rm dex$ for all SMC Cepheids (see discussion in Sect. \ref{sec:SMC_systematics}).  \\

\subsection{Distances}
\label{sect:dist}

Distances to Milky Way Cepheids are taken from the \citet{BailerJones2021} catalog based on the \textit{Gaia} EDR3 parallaxes \citep{GaiaDR3contents}, as well as the associated uncertainties: we adopted the photogeometric distances which are derived using the $(BP - RP)$ color, $G$ magnitude, \textit{Gaia} EDR3 parallaxes, and a direction-dependent prior accounting for the distribution of stellar distances along a line of sight and for interstellar extinction. These distances include the \citet{Lindegren2021_plx_bias} zero-point correction on \textit{Gaia} parallaxes. \citet{Lindegren2021_plx_bias} recommends to include an error of a few $\mu \rm as$ on the parallax zero-point so we assumed a $5\, \rm \mu as$ uncertainty, which is equivalent to including a systematic error of $\sim 0.02 \,\rm mag$ in terms of distance modulus for this sample of MW Cepheids.

Milky Way Cepheids are much brighter than most stars in the Gaia catalogue with visual magnitudes of 6-9 mag, brighter than the range where the Gaia zero-point is well-calibrated.  As a result it is recommended by the \textit{Gaia} team to derive the zero-point in this magnitude range independently from the luminosity using the PL relation.
Following this procedure, \citet{Riess2021} found that the \citet{Lindegren2021_plx_bias} zero-point is overestimated by approximately $14\, \rm \mu as$, which was confirmed by \citet{Zinn2021} from asteroseismology of bright red giants, independently from the PL relation. We therefore applied a small additional correction ($dr$) to the \citet{BailerJones2021} distances in order to take into account this zero-point. A good approximation\footnote{The best way to include the additional 0.014 mas offset would be to infer new corrected distances for all of our MW Cepheids using the \citet{BailerJones2021} method, however it would be beyond the scope of this paper and would give similar results in terms of precision. In particular, running the Bailer-Jones method for such a large number of stars is very time consuming (CPU), therefore it was tested for a few stars only (C. Bailer Jones 2022 private communication, we do not have access to the code to produce these distances). We selected a few Cepheids with low (1\%), typical (3\%) and high (10\%) parallax uncertainties for which the Bailer Jones distances were recomputed after including the additional 0.014 mas offset. The approximated correction reproduces these values to 0.1\% or better, regardless of the parallax precision.}
of this correction is to take $dr = - r^2 \, d\varpi$ where $d\varpi = -0.014\, \rm mas$ and $r$ is the original \citet{BailerJones2021} distance in kpc. Finally, all Cepheids with a Renormalized Unit Weight Error (RUWE) parameter larger than 1.4 were excluded from the sample as likely astrometric binaries.

For LMC Cepheids, we adopt the most precise distance to this galaxy obtained by \citet{Pietrzynski2019} from a sample of 20 detached eclipsing binaries: $49.59 \pm 0.09 \, \rm (stat.) \pm 0.54 \, \rm (syst.) \, kpc$. This corresponds to a full uncertainty of $0.026 \, \rm mag$ in distance modulus. For more precision on the adopted distance, the position of each Cepheid in the LMC is taken into account by applying the planar geometry correction by \citet{Jacyszyn2016} (see \citetalias{Breuval2021}).

With the same technique and assumptions, \citet{Graczyk2020} published the most precise distance to the SMC from 15 eclipsing binary systems distributed around the core: they obtained a distance of $62.44 \pm 0.47 \, \rm (stat.) \pm 0.81 \, \rm (syst.) \, kpc$. This is equivalent to a full uncertainty of $0.032 \, \rm mag$ in distance modulus. To account for the elongated shape of the SMC along the line of sight, we include the geometric model fit to the DEBs and described by the blue lines in Figure 4 of \citet{Graczyk2020}, their equations are provided in \citetalias{Breuval2021} (we also limit the selection of SMC Cepheids to a separation of $< 0.6 \, \rm deg$ between the Cepheids and the SMC center which together with the geometric correction we show greatly reduces their dispersion, see Sect.~\ref{sec:sample_selection} and \ref{sec:comp_lit}). Following \citet{Riess2019}, we include these corrections directly on Cepheid magnitudes. Considering the standard deviation of three different geometry models, \citet{Riess2019} found a systematic uncertainty of $0.002 \, \rm mag$ associated with the LMC geometry: we neglected this contribution to the error budget, since it is widely dominated by other systematics. \\

\section{Method} 
\label{sec:method}

\subsection{Sample selection} 
\label{sec:sample_selection}

Among the Cepheid samples described in the previous section, a selection based on various criteria is performed. First,  only fundamental mode Cepheids are considered: in the Milky Way, the pulsation modes are taken from the new \textit{Gaia} DR3 reclassification by \citet{Ripepi2022Gaia} (see their Table 6). First overtone and mixed-mode pulsators were discarded. A second selection is performed based on the number of epochs available for a given light curve. For the MW sample, only Cepheids with at least 8 data points are considered. Regarding the LMC and SMC samples, a large number of Cepheids have less than 8 measurements per light curve in the NIR and excluding them would drastically reduce the sample, therefore a limit of 5 epochs per star is adopted. For all Cepheids, a minimum uncertainty of 10\% on mean magnitudes is adopted as a precision limit. Additionally, due to the non-negligible depth of the Magellanic Clouds (especially that of the SMC), Cepheids outside of a radius of $3^{\circ}$ around the LMC center and $0.6^{\circ}$ around the SMC center are excluded from the analysis. These regions are found to be optimal as they minimize the scatter of the PL relation (see Sect. \ref{sec:comp_lit}) and together with the geometric correction reduces any potential separation from the mean of the DEBs. 

Finally, a break in the PL/PW relations was identified in the SMC both in the optical and the NIR: the position of this break was found around $\log P \sim 0.4$ by \citet{Udalski1999}, \citet{Sharpee2002}, \citet{Sandage2009} and \citet{Soszynski2010}. \citet{Tammann2011} report a break at a larger period around $\log P \sim 0.55$, and more recently \citet{Subramanian2015} and \citet{Ripepi2016} detected a break at $\log P \sim 0.47$. We perform a cut at $\log P = 0.47$ in the SMC due to this non-linearity, but also at $\log P = 0.4$ in the MW and in the LMC and at $\log P = 2$ in the three galaxies, which allows to prevent for undesirable effects such as confusion of pulsation modes and possible (although not yet detected) breaks at shorter or longer periods. Since the MW and LMC were found to be linear \citep{Inno2016, Ripepi2022VMC}, the period cut does not affect the slope of these samples. For example, changing the break period from $\log P \sim 0.4$ to $\log P \sim 0.47$ only changes the LMC slope by a few millimag/dex and does not impact the results of our analysis. \\

\subsection{Width of the Instability Strip}

The finite width of the instability introduces additional scatter in the PL relation and should be included quadratically as an uncertainty in apparent magnitudes. In \citet{Riess2019}, the width of the instability strip is obtained by taking the scatter of the PL relation ($0.075 \, \rm mag$ from their Table 3), and by subtracting quadratically the errors on photometric measurements (e.g. photometric inhomogeneities, phase corrections) which are of $0.030 \, \rm mag$. They obtained an intrinsic width of $0.069 \, \rm mag$ for the instability strip in the $W_H$ index. Similarly, the values in the $V$ and $I$ bands are $0.22 \, \rm mag$ and $0.14\, \rm mag$ \citep{Macri2006}. In the $J$, $H$ and $K$ bands, the study by \citet{Persson2004} gives a width of $0.11 \, \rm mag$, $0.09 \, \rm mag$ and $0.07\, \rm mag$ for the NIR instability strip. For \textit{Spitzer} bands, we adopted a width of $0.07 \, \rm mag$ \citep{Scowcroft2011, Monson2012}. In the \textit{Gaia} bands $G$, $BP$ and $RP$ as well as in the \textit{Gaia} Wesenheit $W_G$ we adopt for the width of the instability strip the PL dispersion obtained in the LMC by \citet{Ripepi2019} (see their Table 1): $0.19 \, \rm mag$, $0.23 \, \rm mag$, $0.16 \, \rm mag$ and $0.10 \, \rm mag$ respectively. Although their magnitudes are not dereddened, the reddening in the LMC is limited and homogeneous, and the small fraction of highly reddened Cepheids has little impact on the scatter of the PL relation. Indeed, according to \citet{Ripepi2019}, the scatter decreases from $BP$ to $G$ and from $G$ to $RP$, as expected, and is perfectly consistent with the width adopted for the other filters. For the $W_{VI}$ Wesenheit magnitudes we adopt a width of $0.077 \, \rm mag$ from \citet{Soszynski2015}. Finally in $W_{JK}$ and $W_{VK}$ we adopt a width of $0.086 \, \rm mag$ and $0.077 \, \rm mag$ respectively, from the study by \citet{Ripepi2022} that gives a scatter of $0.088 \, \rm mag$ and $0.080 \, \rm mag$ respectively, and photometric errors of the order of $0.020 \, \rm mag$.  \\



\begin{figure}[t!]
\centering
\includegraphics[width=7.8cm]{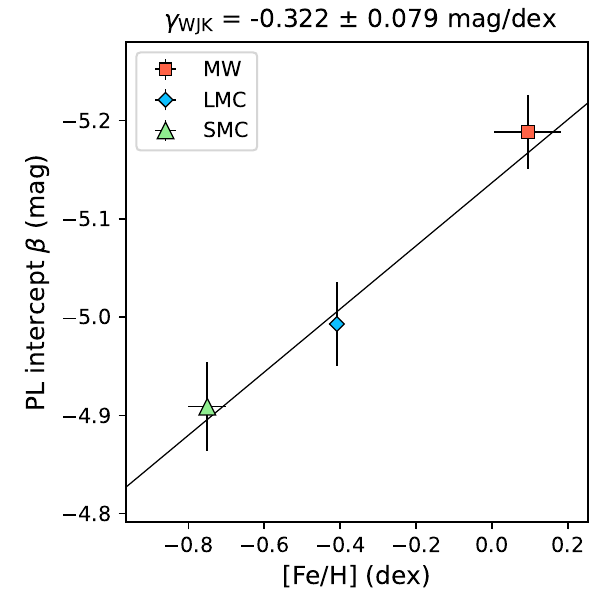} 
\caption{PL intercept ($\beta$) in the $W_{JK}$ Wesenheit index fitted with a common slope in the Milky Way, LMC and SMC as a function of the mean metallicity ([Fe/H]) of the galaxy (Eq. \ref{eq:intercepts_FeH}).}
\label{fig:intercepts_FeH_K}
\end{figure}

\subsection{Period-luminosity-metallicity relation} 
\label{sec:PLZrelation}

The absolute magnitude $M_{\lambda}$ of a star is derived from its apparent magnitude $m_{\lambda}$, its reddening $E(B-V)$ and its distance $d$ in kpc by the equation:
\begin{equation}
M_{\lambda} = m_{\lambda} - R_{\lambda} E(B-V) - 5 \log d - 10
\end{equation}
Apparent magnitudes are corrected for extinction using the standard reddening law from \citet{Fitzpatrick1999} for our $G$, $BP$, $RP$, $V$, $I$, $J$, $H$, $K_S$ magnitudes and the reddening law from \citet{Indebetouw2005} for \textit{Spitzer} filters. We set the $R_V$ parameter to $3.1 \pm 0.1$ which yields the $R_{\lambda}$ values listed in Table \ref{table:filters}. The uncertainty of 0.1 in $R_V$ and its propagation to other filters is intended to characterize the line-of-sight dispersion seen for similar stellar populations (see discussion in Sect. \ref{sec:discussion_rv}). Five Wesenheit indices are also considered \citep{Madore1982} based on a combination of optical and NIR filters, such that $W(\lambda_1, \lambda_2, \lambda_3) = m_{\lambda_1} - R \, (m_{\lambda_2} - m_{\lambda_3})$ with $R = R_{\lambda_1} / (R_{\lambda_2} - R_{\lambda_3})$. For the HST Wesenheit $W_H$ we adopt $R = 0.386$ from \citet{Riess2022} and for the \textit{Gaia} Wesenheit index $W_G$ we use $R = 1.90$ for consistency with \citet{Ripepi2022Gaia}.

To measure the metallicity effect $\gamma$ between the Galactic, LMC and SMC samples, the PL relation of the form
\begin{equation}
M = \alpha \, (\log P - \log P_0) + \beta
\end{equation}
is first fitted in each of the three galaxies, with a common slope $\alpha$ fixed to that of the LMC (since it has the largest number of Cepheids, the lowest PL dispersion and the slope least affected by non-uniformity of individual Cepheid distances). The three PL intercepts $\beta$ are obtained from Monte Carlo sampling of the data and error distributions with 10,000 iterations: to ensure the robustness of the fit, the apparent magnitudes, distances, $R_V$ and $E(B-V)$ values are free to vary within the uncertainties during each iteration. In the case of the MW sample, the $E(B-V)$ values are selected randomly for each star among the three previously described measures of extinction (see Table \ref{table:photom_ZP}) which will naturally account for their covariance. Systematic uncertainties due to photometric zero-points (see Sect. \ref{sec:photometry}) and to distance measurements (see Sect. \ref{sect:dist}) are included quadratically to the intercept errors in the three galaxies. Finally, the metallicity term $\gamma$ of the PL relation as defined in Eq. \ref{eq:PLZ} is obtained by fitting (again with  Monte Carlo sampling) the relation: 
\begin{equation}
\label{eq:intercepts_FeH}
\beta = \gamma \, \rm [Fe/H] + \delta
\end{equation}
where $\beta$ is the PL intercept, [Fe/H] is the mean metallicity in each of the three galaxies and $\delta$ is the fiducial luminosity at $\log P=0.7$ and solar metallicity. As an example, Fig. \ref{fig:intercepts_FeH_K} illustrates the linear fit of Eq. \ref{eq:intercepts_FeH} in the $W_{JK}$ band. \\

\begin{table*}[h!]
\footnotesize
\caption{Results of the PL fit of the form $M = \alpha \,(\log P - 0.7) + \beta$ in the Milky Way, LMC and SMC. In column 3 and 4, $\alpha_{\rm free}$ and $\beta_{\rm free}$ are obtained when both coefficients are free parameters. In columns 5 and 6, $\beta$ is the intercept obtained with the slope $\alpha_{\rm fixed}$ fixed to that of the LMC.}
\centering
\begin{tabular}{c c | c c | c c | c c}
\multicolumn{7}{c}{}  \\
\hline
\hline
Filter & Galaxy & $\alpha_{\rm free}$ & $\beta_{\rm free}$ & $\alpha_{\rm fixed}$ & $\beta$ &  $\sigma$ & $\rm N_{stars}$   \\
\hline
          & MW  & $-2.888 \pm 0.069$ & $-3.197 \pm 0.069$ & $-2.634 \pm 0.030$ & $-3.236 \pm 0.075$ & 0.56 & 445 \\ 
$BP$ & LMC & $-2.634 \pm 0.030$ & $-3.063 \pm 0.013$ & $-2.634 \pm 0.030$ & $-3.063 \pm 0.029$ & 0.23 & 1593 \\
          & SMC & $-2.523 \pm 0.070$ & $-2.961 \pm 0.018$ & $-2.634 \pm 0.030$ & $-2.957 \pm 0.037$ & $0.29$ & 287 \\ 
\hline
        & MW  & $-2.631 \pm 0.087$ & $-3.356 \pm 0.049$ & $-2.715 \pm 0.029$ & $-3.338 \pm 0.060$ & $0.22$ & 183 \\ 
$V$ & LMC & $-2.715 \pm 0.029$ & $-3.191 \pm 0.011$ & $-2.715 \pm 0.029$ & $-3.191 \pm 0.035$ & $0.21$ & 1609  \\
       & SMC & $-2.596 \pm 0.067$ & $-3.079 \pm 0.018$ & $-2.715 \pm 0.029$ & $-3.074 \pm 0.041$ & $0.28$ & 291 \\ 
\hline
        & MW  & $-3.112 \pm 0.060$ & $-3.578 \pm 0.058$ & $-2.816 \pm 0.025$ & $-3.623 \pm 0.064$ & $0.49$ & 446 \\ 
$G$ & LMC & $-2.816 \pm 0.025$ & $-3.335 \pm 0.010$ & $-2.816 \pm 0.025$ & $-3.335 \pm 0.028$ & $0.18$ & 1602 \\ 
        & SMC & $-2.758 \pm 0.059$ & $-3.221 \pm 0.016$ & $-2.816 \pm 0.025$ & $-3.218 \pm 0.036$ & $0.25$ & 288 \\ 
\hline
          & MW  & $-3.037 \pm 0.046$ & $-3.850 \pm 0.042$ & $-2.918 \pm 0.019$ & $-3.868 \pm 0.047$ & $0.44$ & 446 \\ 
$RP$ & LMC & $-2.918 \pm 0.019$ & $-3.812 \pm 0.008$ & $-2.918 \pm 0.019$ & $-3.813 \pm 0.027$ & $0.16$ & 1584 \\ 
          & SMC & $-2.853 \pm 0.044$ & $-3.722 \pm 0.012$ & $-2.918 \pm 0.019$ & $-3.718 \pm 0.034$ & $0.22$ & 287 \\
\hline
      & MW  & $-2.858 \pm 0.063$ & $-4.005 \pm 0.031$ & $-2.950 \pm 0.018$ & $-3.988 \pm 0.043$ & $0.19$ & 157 \\ 
$I$ & LMC & $-2.950 \pm 0.018$ & $-3.851 \pm 0.007$ & $-2.950 \pm 0.018$ & $-3.851 \pm 0.034$ & $0.14$ & 1687 \\ 
      & SMC & $-2.971 \pm 0.040$ & $-3.772 \pm 0.010$ & $-2.950 \pm 0.018$ & $-3.770 \pm 0.039$ & $0.21$ & 300 \\ 
\hline
       & MW  & $-3.207 \pm 0.075$ & $-4.525 \pm 0.028$ & $-3.097 \pm 0.013$ & $-4.548 \pm 0.036$ & $0.19$ & 71 \\ 
$J$ & LMC & $-3.097 \pm 0.013$ & $-4.335 \pm 0.004$ & $-3.097 \pm 0.013$ & $-4.335 \pm 0.038$ & $0.13$ & 1644 \\ 
       & SMC & $-3.001 \pm 0.026$ & $-4.292 \pm 0.007$ & $-3.097 \pm 0.013$ & $-4.287 \pm 0.040$ & $0.17$ & 299  \\ 
\hline
        & MW  & $-3.296 \pm 0.066$ & $-4.787 \pm 0.022$ & $-3.161 \pm 0.013$ & $-4.816 \pm 0.033$ & $0.18$ & 70 \\ 
$H$ & LMC & $-3.161 \pm 0.013$ & $-4.677 \pm 0.004$ & $-3.161 \pm 0.013$ & $-4.677 \pm 0.037$ & $0.09$ & 751 \\ 
        & SMC & $-3.207 \pm 0.023$ & $-4.578 \pm 0.006$ & $-3.161 \pm 0.013$ & $-4.581 \pm 0.039$ & $0.17$ & 290  \\ 
\hline
       & MW  & $-3.235 \pm 0.065$ & $-4.929 \pm 0.021$ & $-3.222 \pm 0.008$ & $-4.932 \pm 0.033$ & $0.17$ & 65 \\ 
$K$ & LMC & $-3.222 \pm 0.008$ & $-4.713 \pm 0.002$ & $-3.222 \pm 0.008$ & $-4.713 \pm 0.033$ & $0.09$ & 1653  \\ 
       & SMC & $-3.198 \pm 0.017$ & $-4.649 \pm 0.004$ & $-3.223 \pm 0.008$ & $-4.647 \pm 0.039$ & $0.15$ & 299 \\ 
\hline
                                 & MW  & $-3.457 \pm 0.076$ & $-4.922 \pm 0.037$ & $-3.324 \pm 0.038$ & $-4.968 \pm 0.032$ & $0.20$ & 21  \\
$[3.6 \, \rm \mu m]$ & LMC & $-3.324 \pm 0.038$ & $-4.791 \pm 0.024$ & $-3.324 \pm 0.038$ & $-4.791 \pm 0.027$ & $0.11$ & 66 \\ 
                                & SMC & $-3.615 \pm 0.063$ & $-4.539 \pm 0.038$ & $-3.324 \pm 0.038$ & $-4.703 \pm 0.036$ & $0.12$ & 23 \\ 
\hline
                                & MW  & $-3.365 \pm 0.074$ & $-4.913 \pm 0.037$ & $-3.233 \pm 0.040$ & $-4.961 \pm 0.034$ & $0.21$ & 21 \\ 
$[4.5 \, \rm \mu m]$ & LMC & $-3.233 \pm 0.040$ & $-4.821 \pm 0.024$ & $-3.233 \pm 0.040$ & $-4.821 \pm 0.027$ & $0.11$ & 66 \\ 
                                & SMC & $-3.582 \pm 0.064$ & $-4.574 \pm 0.037$ & $-3.233 \pm 0.040$ & $-4.770 \pm 0.035$ & $0.14$ & 23 \\ 
\hline
             & MW  & $-3.422 \pm 0.026$ & $-4.995 \pm 0.007$ & $-3.338 \pm 0.012$ & $-5.005 \pm 0.021$ & $0.32$ & 596 \\
$W_G$ & LMC & $-3.338 \pm 0.012$ & $-4.791 \pm 0.003$ & $-3.338 \pm 0.012$ & $-4.791 \pm 0.026$ & $0.11$ & 1591 \\ 
             & SMC & $-3.388 \pm 0.025$ & $-4.683 \pm 0.006$ & $-3.338 \pm 0.012$ & $-4.686 \pm 0.033$ & $0.14$ & 286  \\ 
\hline
                 & MW  & $-3.197 \pm 0.036$ & $-4.913 \pm 0.011$ & $-3.291 \pm 0.010$ & $-4.895 \pm 0.038$ & $0.18$ & 157 \\ 
$W_{VI}$ & LMC & $-3.291 \pm 0.010$ & $-4.771 \pm 0.002$ & $-3.291 \pm 0.010$ & $-4.771 \pm 0.038$ & $0.09$ & 1606  \\ 
                & SMC & $-3.317 \pm 0.021$ & $-4.726 \pm 0.005$ & $-3.291 \pm 0.010$ & $-4.728 \pm 0.043$ & $0.13$ & 288  \\ 
\hline
                 & MW  & $-3.345 \pm 0.064$ & $-5.182 \pm 0.022$ & $-3.323 \pm 0.009$ & $-5.188 \pm 0.038$ & $0.18$ & 63  \\ 
$W_{JK}$ & LMC & $-3.323 \pm 0.009$ & $-4.993 \pm 0.002$ & $-3.323 \pm 0.009$ & $-4.993 \pm 0.042$ & $0.09$ & 1653  \\ 
                 & SMC & $-3.350 \pm 0.018$ & $-4.907 \pm 0.005$ & $-3.323 \pm 0.009$ & $-4.909 \pm 0.045$ & $0.13$ & 298  \\ 
\hline
                  & MW  & $-3.246 \pm 0.068$ & $-5.149 \pm 0.021$ & $-3.255 \pm 0.011$ & $-5.146 \pm 0.039$ & $0.21$ & 57 \\ 
$W_{VK}$ & LMC & $-3.255 \pm 0.011$ & $-4.905 \pm 0.002$ & $-3.255 \pm 0.011$ & $-4.905 \pm 0.038$ & $0.08$ & 1500 \\ 
                  & SMC & $-3.293 \pm 0.018$ & $-4.844 \pm 0.005$ & $-3.255 \pm 0.011$ & $-4.847 \pm 0.044$ & $0.14$ & 288  \\ 
\hline
             & MW  & $-3.361 \pm 0.056$ & $-4.949 \pm 0.019$ & $-3.305 \pm 0.038$ & $-4.964 \pm 0.024$ & $0.16$ & 60 \\
$W_H$ & LMC & $-3.305 \pm 0.038$ & $-4.816 \pm 0.018$ & $-3.305 \pm 0.038$ & $-4.823 \pm 0.027$ & $0.08$ & 70 \\ 
             & SMC & $-$ & $-$ & $-$ & $-$ & $-$ & $-$ \\ 
\hline
\multicolumn{7}{c}{}  \\
\end{tabular}
\label{table:PL}
\end{table*}

\section{Results} 
\label{sec:results}

\subsection{The PL relation}
\label{sec:PL}

\begin{figure}[t!]
\centering
\includegraphics[width=8.4cm]{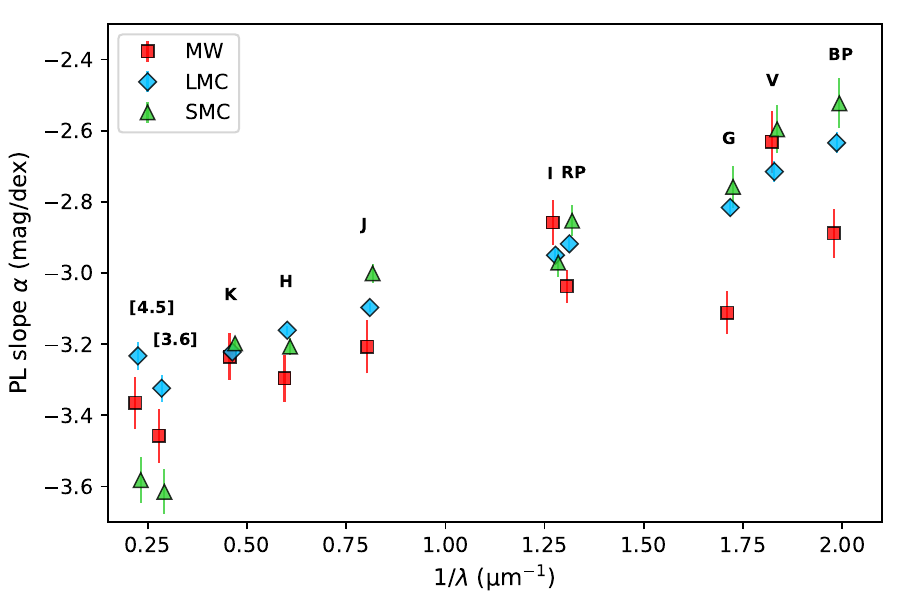} 
\includegraphics[width=8.4cm]{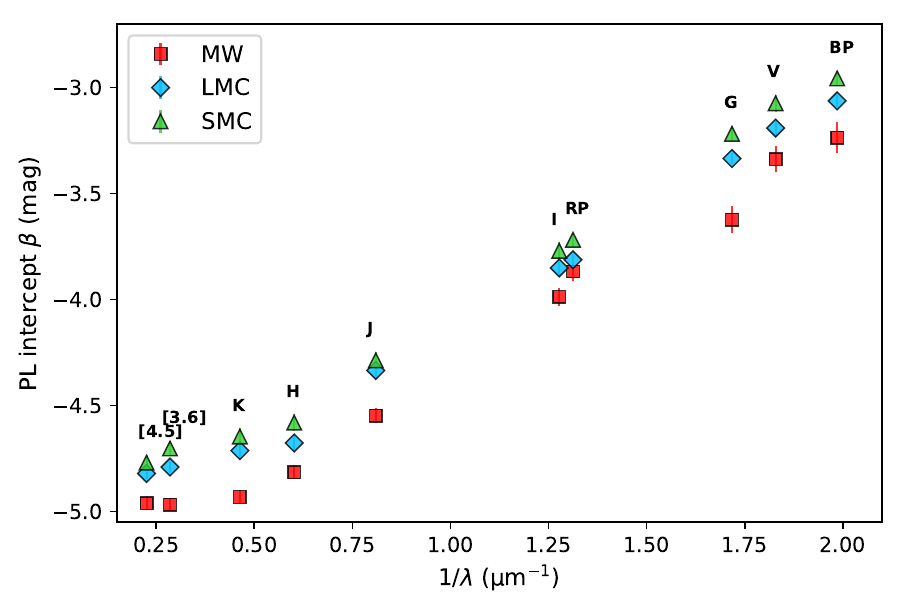} 
\caption{\textbf{Top:} PL slope ($\alpha$) in the Milky Way and Magellanic Clouds represented with the inverse of wavelength. \textbf{Bottom:} PL intercept ($\beta$) obtained in the Milky Way and Magellanic Clouds with a slope fixed to that of the LMC, represented with the inverse of wavelength.} 
\label{fig:alpha_beta_lambda}
\end{figure}

In Sect. \ref{sec:PLZ}, the metallicity term ($\gamma$) of the PL relation will be derived when the slopes ($\alpha$) are fixed to the same value in the three galaxies, in order to directly compare the intercepts ($\beta$). However in this section we first calibrate the PL relation of the form $M = \alpha (\log P -0.7) + \beta$ in each galaxy where both the slope and the intercept are free to vary. This allows to check the consistency of the slopes in the three galaxies. The coefficients are listed in Table \ref{table:PL} and represented in Fig. \ref{fig:alpha_beta_lambda}. 

Generally, the PL slope obtained for the LMC sample agrees to better than $3\sigma$ with that of the MW and SMC samples. The only exceptions are the $G$ and $BP$ bands where the MW and LMC slopes differ by $4\sigma$ and $3 \sigma$ respectively, both \textit{Spitzer} bands where the LMC slope is shallower than in the SMC by $4 \sigma$, and the $J$ filter where the LMC and SMC PL slopes differ by $3.3 \sigma$. However, we see no strong evidence to reject the hypothesis of a common slope in the three galaxies (see Sect. \ref{sec:PLZ}): the disagreement between the LMC and the SMC slopes in both \textit{Spitzer} filters can be traced back to the strict selection of the core region of the SMC ($R<0.6$ deg), which only leaves 22 out of the 90 original Cepheids. When including all 90 Cepheids from the SMC sample, we can closely reproduce the slopes reported by \citet{Scowcroft2016}. In Sect. \ref{sec:gamma_lambda} we discuss the impact of the SMC sample selection on the values of $\gamma$. In $J$, adopting the SMC slope instead of the LMC slope changes the $\gamma$ parameter by only 0.014 mag/dex which is negligible compared to the uncertainties. Finally, adopting the MW slope in \textit{Gaia} filters changes the gamma term by 0.013 mag at most.

We also note that our PL slopes in the 3 galaxies are in excellent agreement with those reported by \citet{Ripepi2022Gaia} in the $G$ and $W_G$ filters. Our slopes agree well with \citet{Subramanian2015} in $V$ and $I$, and with \citet{Ripepi2016} in $J$, $K$, $W_{JK}$ and $W_{VK}$ for the SMC sample.

As expected, the dispersion of the PL relation decreases from the optical to the infrared, which is a consequence of the sensitivity of each filter to the extinction and of the width of the instability strip. In the Milky Way as well as in both Magellanic Clouds, the PL slope $\alpha$ generally becomes steeper and the PL intercept $\beta$ becomes more negative (i.e. brighter) from the optical towards the infrared. 
Due to the presence of a large CO absorption band aligned with the $[4.5 \, \rm \mu m]$ filter \citep{Marengo2010, Freedman2011b, Scowcroft2011}, in section \ref{sec:gamma_lambda} the $[4.5 \, \rm \mu m]$ filter is ignored in the fit of the $\gamma = f(\lambda)$ relation. In the Wesenheit $W_H$ band we obtain a slope of $-3.305 \pm 0.038 \, \rm mag/dex$ in the LMC which is fully compatible with the slope of $-3.299 \pm 0.015 \, \rm mag/dex$ derived by the SH0ES team \citep{Riess2022}.

In the Milky Way, the PL relation generally shows a larger dispersion than in the Magellanic Clouds because of the higher extinction and non-uniform distances. The PL relations in Wesenheit indices show a low dispersion, as expected from their insensitivity to extinction. In some filters only a small number of Cepheids is listed in Table \ref{table:PL}: this is due to the various selection criteria applied to the samples, such as the upper limit of 1.4 on the RUWE parameter, the limited radius around the SMC center and the cuts in periods. Finally, for a given filter, the PL intercept in the Milky Way is more negative than in the LMC and even more than in the SMC (see Fig. \ref{fig:alpha_beta_lambda}), indicating a negative sign for the $\gamma$ term.  \\

\subsection{The PLZ relation} 
\label{sec:PLZ}

After fixing the PL slope to that of the LMC (Table \ref{table:PL}) in the three galaxies, we solve for Eq. \ref{eq:intercepts_FeH} with a Monte Carlo sampling where both the intercepts ($\beta$) and the mean [Fe/H] values of each sample are free to vary within their error bars. The $\gamma$ and $\delta$ coefficients obtained for the PLZ relation are listed in Table \ref{table:PLZ}. 

 All $\gamma$ values over a wavelength range of $0.5 - 4.5 \, \rm \mu m$ are negative (with a significance of $2.6 \sigma$ to $7.5 \sigma$), meaning that metal-rich Cepheids are brighter than metal-poor ones. The $\gamma$ values range between a minimum of $-0.178 \pm 0.068 \, \rm mag/dex$ (in $RP$) and a maximum of $-0.462 \pm 0.089 \, \rm mag/dex$ (in $G$) with a dispersion of $0.05 \, \rm mag/dex$. In all filters, the $\gamma$ values are in good agreement with those obtained by \citet{Gieren2018} and \citetalias{Breuval2021}, especially in the NIR, but significantly stronger than the effect detected by \citet{Wielgorski2017}, which was close to zero (see discussion in Sect. \ref{sec:comp_lit}). The metallicity effect in \textit{Gaia} filters is similar to that in ground optical filters ($V$, $I$), however the $G$ band and $W_G$ Wesenheit index show a stronger effect (see discussion in Sect. \ref{sec:discussion_gaia}). 

\begin{table}[t!]
\small
\caption{Results of the fit of the form $\beta = \gamma \, \rm [Fe/H] + \delta $ (Eq. \ref{eq:intercepts_FeH}) obtained from a comparison of the PL intercepts ($\beta$) in the Milky Way, LMC and SMC. The slope $\alpha$ is fixed to that of the LMC sample (see Table \ref{table:PL}).}
\centering
\begin{tabular}{c c c c c c}
\hline
\hline
Filter & $\gamma$ & $\sigma$  & $\delta$ & $\sigma$ & $\rm N_{stars}$   \\
\hline
$BP$ & $-0.320$ & $0.095$ & $-3.194$ & $0.050$ & 2325 \\
 $V$ & $-0.311$ & $0.082$ & $-3.314$ & $0.046$ & 2083 \\
 $G$ & $-0.462$ & $0.089$ & $-3.539$ & $0.047$ & 2336 \\
 $RP$ & $-0.178$ & $0.068$ & $-3.873$ & $0.036$ & 2317 \\
 $I$ & $-0.247$ & $0.068$ & $-3.956$ & $0.038$ & 2144 \\
 $J$ & $-0.294$ & $0.066$ & $-4.478$ & $0.037$ & 2014 \\
 $H$ & $-0.275$ & $0.065$ & $-4.789$ & $0.036$ & 1111 \\
 $K$ & $-0.321$ & $0.068$ & $-4.860$ & $0.034$ & 2017 \\
$[3.6\, \rm \mu m]$ & $-0.292$ & $0.057$ & $-4.915$ & $0.031$ & 110 \\
$[4.5\, \rm \mu m]$ & $-0.204$ & $0.057$ & $-4.911$ & $0.029$ & 110 \\
\hline
$W_G$    & $-0.384$ & $0.051$ & $-4.958$ & $0.025$ & 2473 \\
$W_{VI}$ & $-0.201$ & $0.071$ & $-4.864$ & $0.035$ & 2051 \\
$W_{JK}$ & $-0.322$ & $0.079$ & $-5.137$ & $0.042$ & 2014 \\
$W_{VK}$ & $-0.332$ & $0.081$ & $-5.066$ & $0.042$ & 1845 \\
\, \, \, $W_H$ $^{(*)}$ & $-0.280$ & $0.078$ & $-4.939$ & $0.027$ & 130 \\
\hline
\end{tabular}
{\flushleft \textbf{Note:} ${(*)}$ Does not include SMC sample (no HST photometry) or individual metallicities in the MW or Cepheids in SN Ia hosts as used in \citet{Riess2022}.  \par}
\label{table:PLZ}
\end{table}


We tested the hypothesis of a common slope in the three galaxies by fixing the slope to that of the SMC instead of that of the LMC: we obtained similar $\gamma$ values at the $0.8 \sigma$ level or better, confirming the validity of our hypothesis. \\

\section{Discussion} 
\label{sec:discussion}

\subsection{Potential issues with the \textit{Gaia} Wesenheit index}
\label{sec:discussion_gaia}

\begin{figure*}[t!]
\centering
\includegraphics[width=14.0cm]{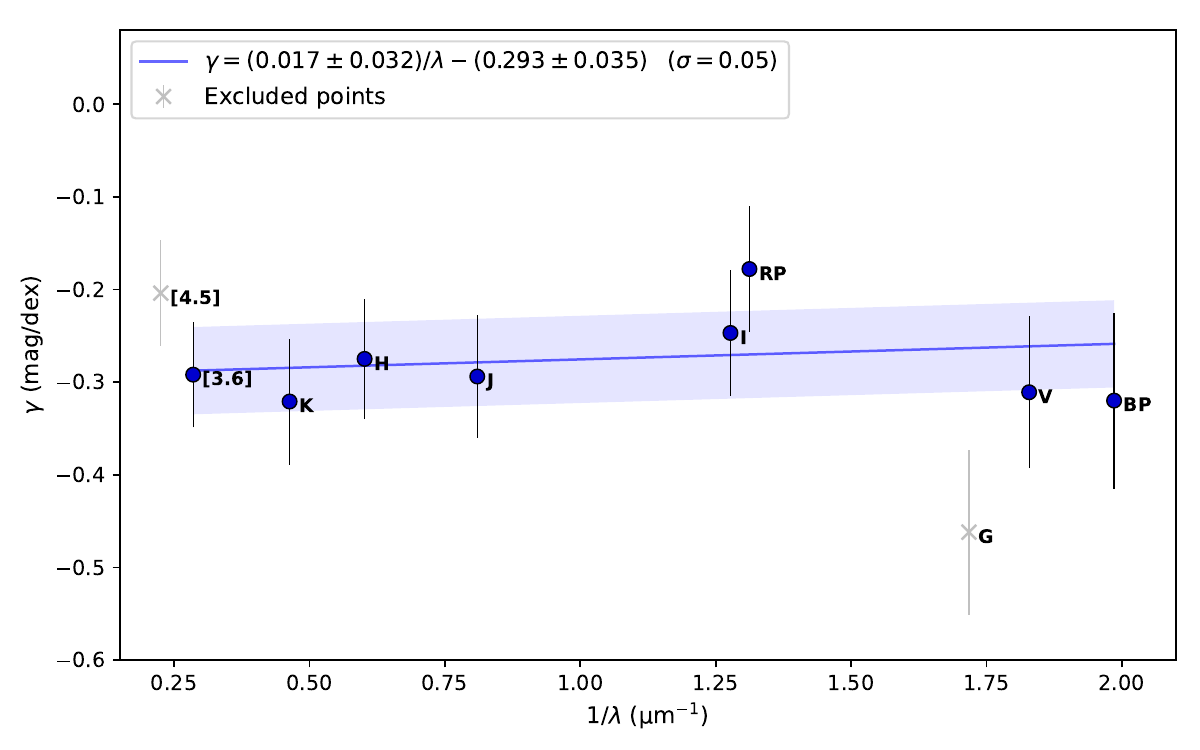} 
\caption{Metallicity effect ($\gamma$) as a function of the inverse of wavelength ($1/\lambda$). }
\label{fig:gamma_lambda}
\end{figure*}

In the \textit{Gaia} Wesenheit index $W_G$, we derive a strong effect of $-0.384 \pm 0.051 \, \rm mag/dex$, slightly shallower but still close to the previous result by \citet{Ripepi2022} who obtained $-0.520 \pm 0.090 \, \rm mag/dex$ from Milky Way Cepheids. The metallicity effect in this Wesenheit index is surprisingly strong compared to other filters or other Wesenheit indices, but is comparable to that in the \textit{Gaia} $G$ band ($-0.462 \pm 0.089 \, \rm mag/dex$). This could be explained by the particularly large width of the $G$ filter (almost 800 nm) 
and suggests that the results based on \textit{Gaia} $G$ band photometry should be treated particularly carefully. For these reasons, the $G$ band is ignored in the fit of the relation between $\gamma$ and $\lambda$ in Fig. \ref{fig:gamma_lambda}. Additionally, Wesenheit indices have been established to minimize the effects of interstellar extinction, however they are not totally independent of the reddening law since they rely on the $R$ coefficient (see Sect. \ref{sec:method}).  \\

\subsection{A relationship between $\gamma$ and $\lambda$}
\label{sec:gamma_lambda}

As the filters used in this analysis cover a large wavelength range, we can measure a dependence between the metallicity term $\gamma$ and the wavelength. When fitting a linear relationship between $\gamma$ and $1/\lambda$ through the points of Fig.~\ref{fig:gamma_lambda} after excluding the $[4.5 \, \rm \mu m]$ filter (see Sect. \ref{sec:PL}) and the $G$ band (see Sect. \ref{sec:discussion_gaia}), we derive the following relation:
\begin{equation}
\label{eq:gamma_lam}
\gamma = \frac{0.017 \pm 0.032}{\lambda} - (0.293 \pm 0.035) \, \, \,  \rm mag/dex
\end{equation}
with $\sigma = 0.05$ mag/dex. The slope of Eq. \ref{eq:gamma_lam} shows that the metallicity effect is mostly uniform over the wavelength range $0.5 - 4.5 \, \rm \mu m$. Compared to the luminosity dependence it indicates that Cepheid colors are relatively insensitive to metallicity.

To verify this is not related to any use of Cepheid colors in reddening measurements, we repeated the analysis after discarding reddening estimates based on color (i.e. only the reddening maps by \citet{Green2019} are used): we obtain a similar dependence ($\gamma \sim 0.038 \pm 0.043 / \lambda $) which confirms the previous finding.

As mentioned in Sect. \ref{sec:PL}, the PL slope in \textit{Spitzer} bands for the SMC sample depends on the adopted region around the SMC center. The selection corresponding to $R<0.6$ deg excludes a large fraction of the initial sample and returns PL slopes that are more negative than expected. With a more moderate selection of $R<1.2$ deg, the SMC slopes are in better agreement with \citet{Scowcroft2016} and the $\gamma$ values become slightly shallower, with $-0.279 \pm 0.060$ and $-0.194 \pm 0.056$ mag/dex in $[3.6\, \rm \mu m]$ and $[4.5\, \rm \mu m]$ respectively. This would revise Eq. \ref{eq:gamma_lam} to $\gamma = (0.012 \pm 0.032)/\lambda - (0.286 \pm 0.035) \rm \, mag/dex$.  \\

 \subsection{Reddening law}
\label{sec:discussion_rv}

The correction for dust extinction and the assumption of a reddening law are critical steps in the calibration of the distance scale. The parameter $R_V = A_V / E(B-V)$ is related to the average size of dust grains and gives a physical basis for the variations in extinction curves. Although the differences in $R_V$ are relatively small between the Milky Way and Magellanic Clouds, they can still impact the calibration of the Leavitt law. In the Milky Way, most studies are based on the assumption $R_V = 3.1$ \citep{Cardelli1989}, while \citet{Gordon2003} report an average of $R_V = 3.41 \pm 0.06$ in the LMC and $R_V = 2.74 \pm 0.13$ in the SMC. They conclude that LMC and SMC extinction curves are similar qualitatively to those derived in the Milky Way. But even in the Milky Way the extinction curve was shown to be highly spatially variable \citep{Fitzpatrick2019}. 

Assuming a different reddening law or $R_V$ value across different galaxies is possible but more complex \citep[see][Appendix D]{Riess2022}: since the $R$ ratio in Wesenheit indices multiplies a color term, it requires to separate the contribution of the color that results from dust reddening by first subtracting the intrinsic color of Cepheids, which can be done using a period-color relation. However, \citet{Riess2022} concluded that determining individual values of $R$ was not very informative due to large uncertainties on both color and brightness.

In the present work, we adopted the standard reddening law from \citet{Fitzpatrick1999} for our $G$, $BP$, $RP$, $V$, $I$, $J$, $H$, $K_S$ magnitudes and the reddening law from \citet{Indebetouw2005} for \textit{Spitzer} filters with a uniform $R_V$ value of $3.1 \pm 0.1$. We note that the uncertainty on this parameter is usually neglected in most studies, even when combining Cepheid samples in different galaxies \citep[e.g.][]{Wielgorski2017, Gieren2018, Owens2022}. While it is a reasonnable assumption for the Milky Way and LMC, the Small Magellanic Cloud is likely to have a lower $R_V$ value \citep{Gordon2003}, however this value has not been measured for our population of Cepheids, so it is still unclear whether it applies to the present sample. For simplicity and consistency between the three galaxies, we assumed the same $R_V$ in the three samples. For each filter we also included the uncertainties on the $A_{\lambda}/A_V$ ratios by varying $R_V$ by $\pm 0.1$ with the \texttt{dust\_extinction} Python package\footnote{\href{https://dust-extinction.readthedocs.io/en/stable/}{https://dust-extinction.readthedocs.io/en/stable/}}.

While it is not recommended to vary $R_V$ between host galaxies for Wesenheit indices \citep[][appendix D]{Riess2022}, we tested the effect of changing the $R_V$ value to $2.74 \pm 0.13$ in the SMC for single filters only. We find that the metallicity effect becomes stronger in absolute sense (i.e. more negative) by 0.020 to 0.040 mag/dex in optical bands and by at most 0.008 mag/dex in the NIR. These changes are well within the error bars listed in Table \ref{table:PLZ} and result in a shallower dependence between $\gamma$ and wavelength, with Eq. \ref{eq:gamma_lam} becoming $\gamma = (0.005 \pm 0.035)/\lambda - (0.288 \pm 0.039) \rm \, mag/dex$.    \\

\begin{figure*}[h!]
\centering
\includegraphics[width=8.8cm]{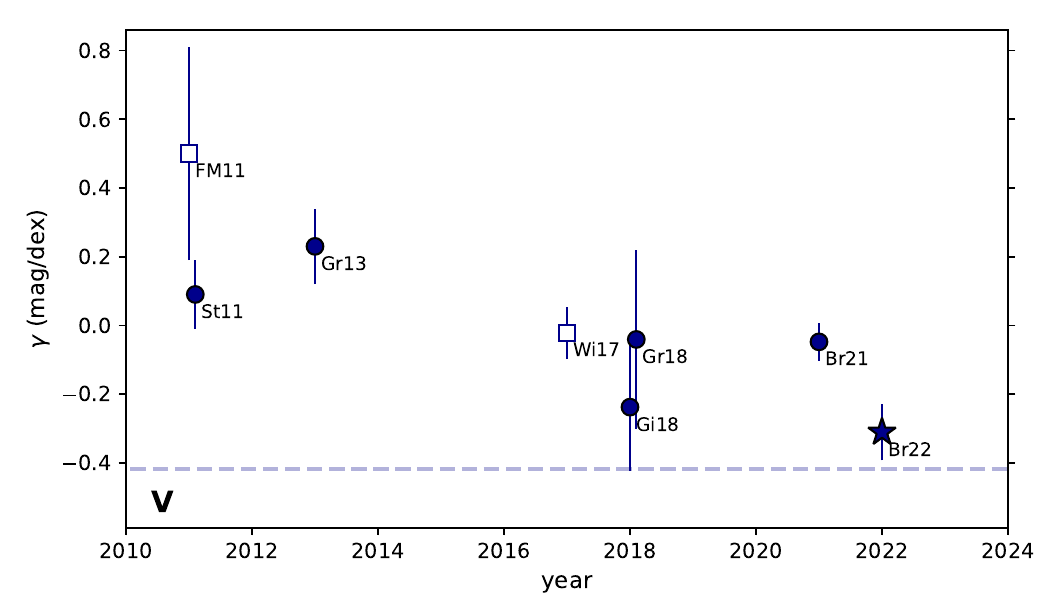} ~
\includegraphics[width=8.8cm]{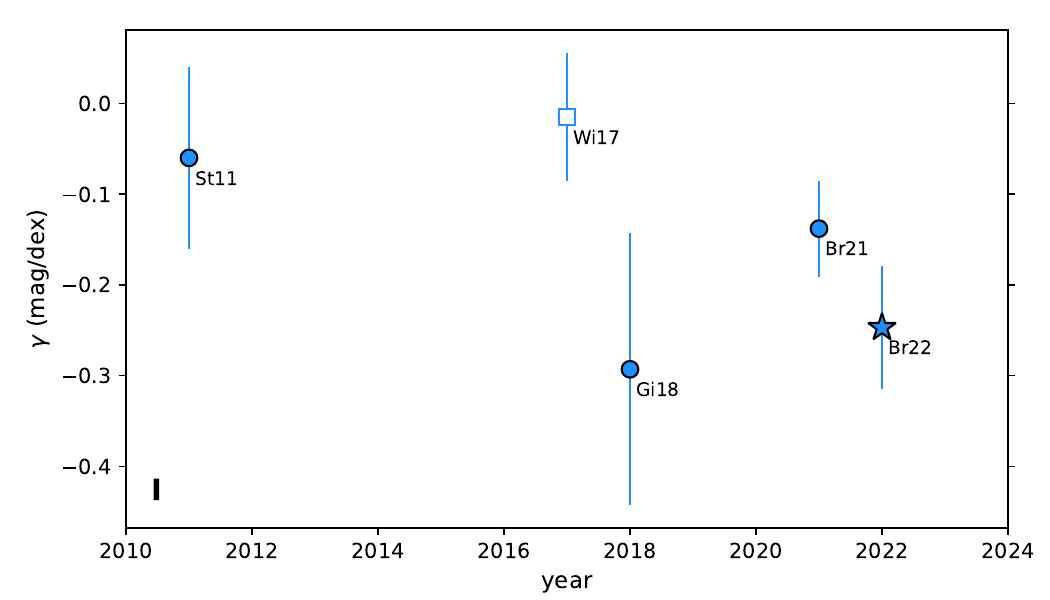} ~
\includegraphics[width=8.8cm]{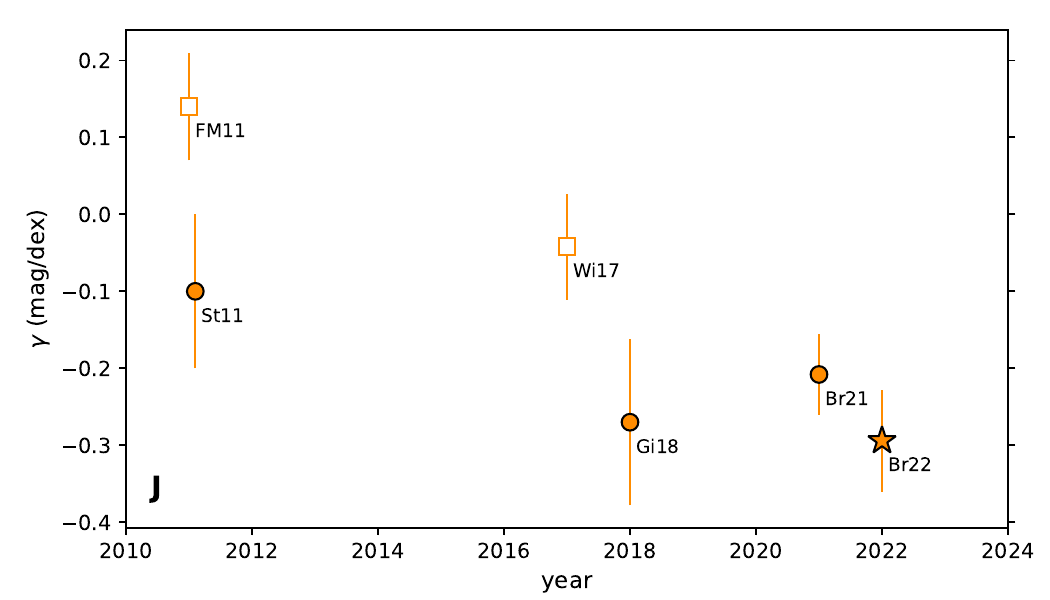} ~
\includegraphics[width=8.8cm]{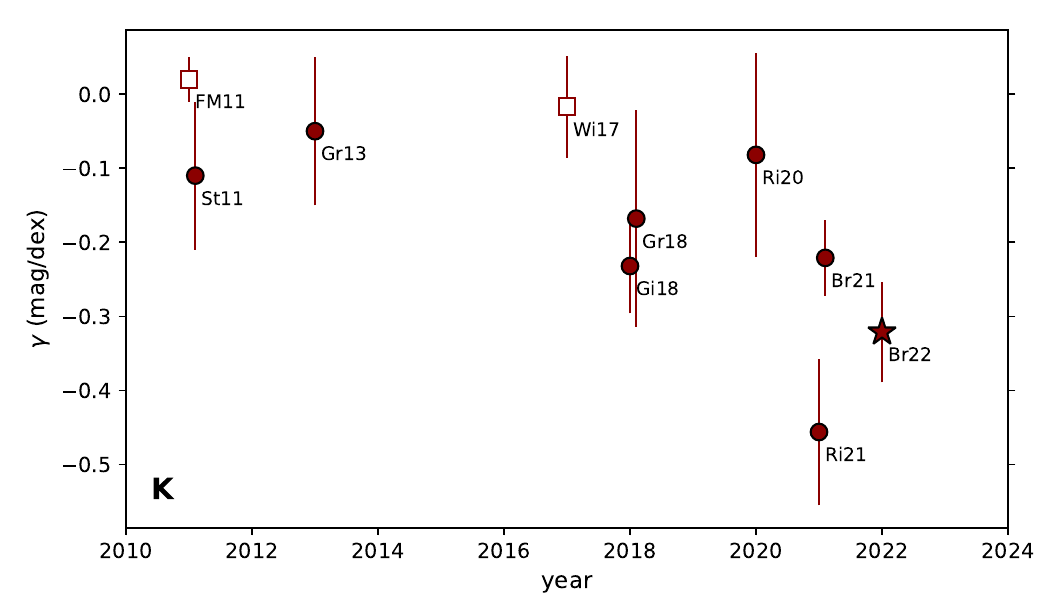} ~
\includegraphics[width=8.8cm]{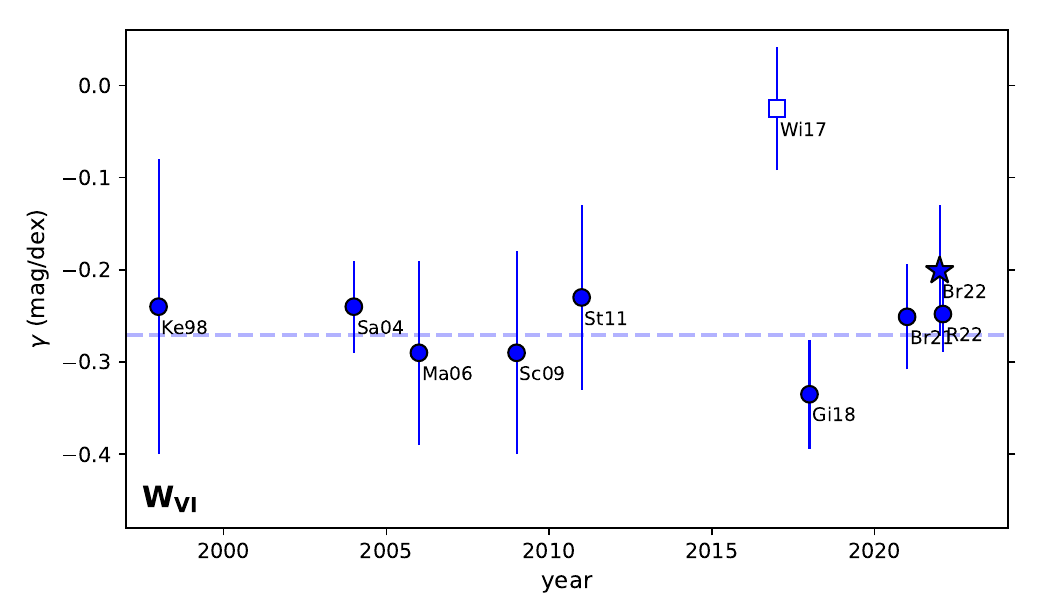} ~
\includegraphics[width=8.8cm]{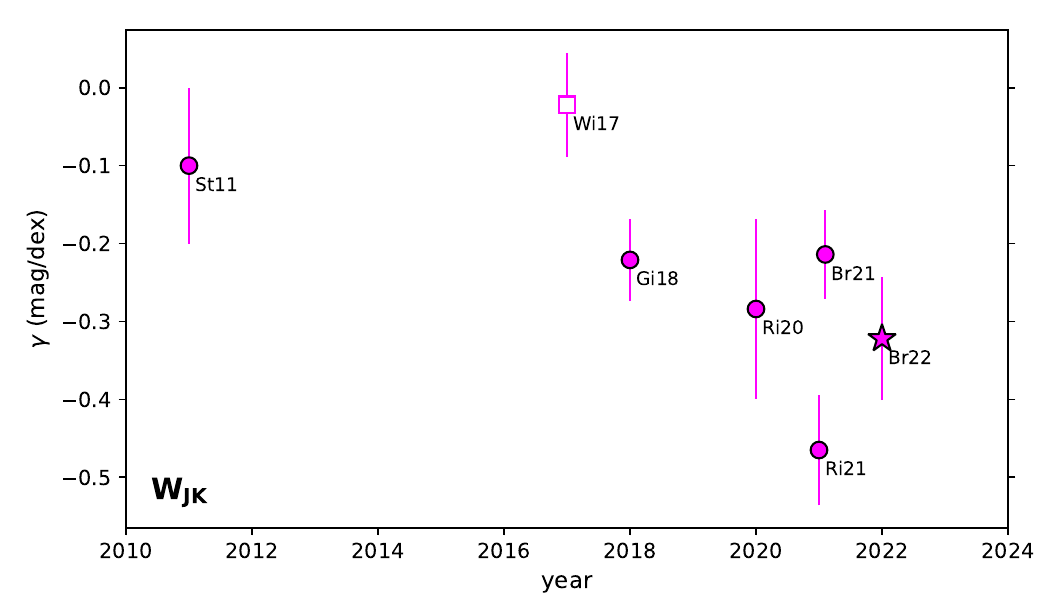} ~
\includegraphics[width=8.8cm]{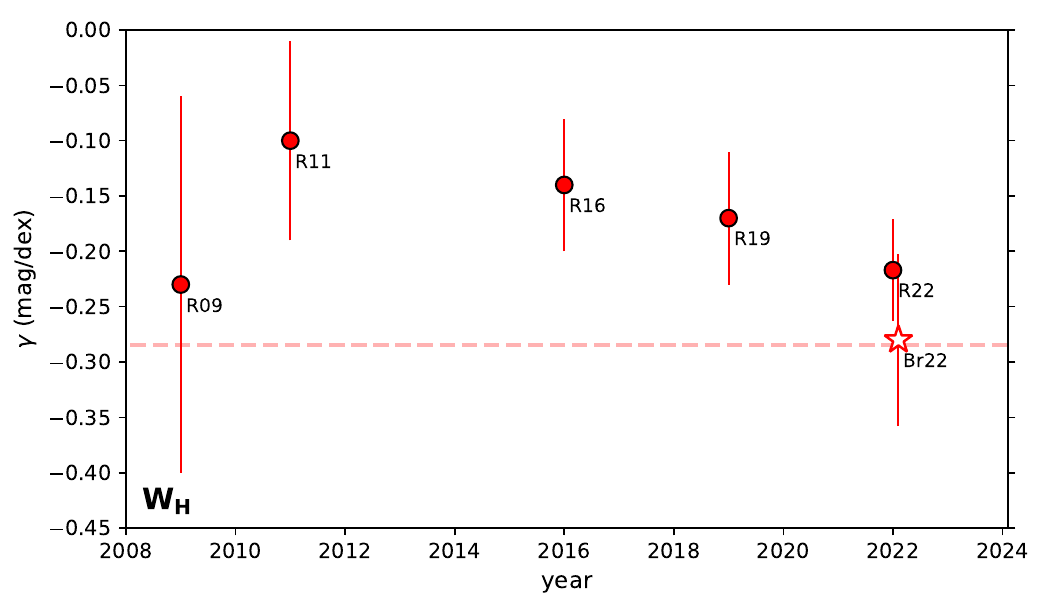} ~
\caption{Evolution of the metallicity term $\gamma$ in different filters over time. Open square symbols indicate the studies in which we identified issues which likely affect the accuracy of the corresponding $\gamma$ values. The open star symbol in the bottom figure indicates that the present work in $W_H$ does not include HST photometry for the SMC sample. \textbf{References}:  (Ke98): \citet{Kennicutt1998}, (Sa04): \citet{Sakai2004}, (Ma06): \citet{Macri2006}, (Sc09): \citet{Scowcroft2009}, (R09): \citet{Riess2009}, (FM11): \citet{Freedman2011}, (St11): \citet{Storm2011b}, (R11): \citet{Riess2011}, (Gr13): \citet{Groenewegen2013}, (R16): \citet{Riess2016}, (Wi17): \citet{Wielgorski2017}, (Gi18): \citet{Gieren2018}, (Gr18): \citet{Groenewegen2018}, (R19): \citet{Riess2019}, (Ri20): \citet{Ripepi2020}, (Ri21): \citet{Ripepi2021}, (R22): \citet{Riess2022}, (Br21): \citet{Breuval2021}, (Br22): present work. The dashed colored line represents the prediction of $\gamma$ from \citet{Anderson2016} using Geneva evolution models including the effects of rotation.  }
\label{fig:gamma_time}
\end{figure*}

\subsection{SMC mean metallicity}
\label{sec:SMC_systematics}

In the \citet{Romaniello2021} reanalysis of the LMC metallicity, a shift of 0.07 dex was detected compared with the previous value by \citet{Romaniello2008}. This offset is due to a difference in temperature in the abundance analysis. In light of this reanalysis, we can reasonably expect a similar shift in the SMC mean metallicity: following the same procedure, it is plausible that a reanalysis of the original SMC data leads to a revised value of about [Fe/H] = $-0.90 \pm 0.05$ dex. While a more detailed analysis is required before adopting this value as our final SMC metallicity, we can easily measure the impact that this change would have on the $\gamma$ values. Replacing the original SMC metallicity of $-0.75$ dex by a more metal-poor value of $-0.90$ dex gives shallower values of the metallicity effect. Typically, the gamma values change by $0.020$ mag/dex (e.g. in $RP$) to $0.055$ mag/dex (e.g. in $JHK$), which is comprised within the error bars. 
Overall, this change in the SMC metallicity results in a shallower dependence between $\gamma$ and the effective wavelength, with: $\gamma = (0.007 \pm 0.024)/\lambda - (0.235 \pm 0.026)$ mag/dex (see Fig. \ref{fig:gamma_lambda_metalpoor} in Appendix).

\begin{figure*}[t!]
\centering
\includegraphics[width=8.8cm]{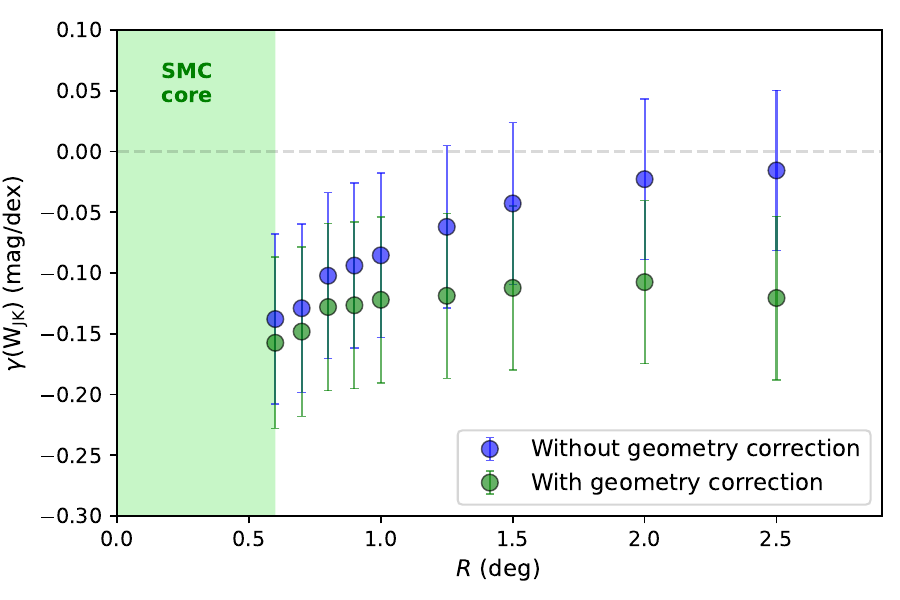} ~
\includegraphics[width=8.8cm]{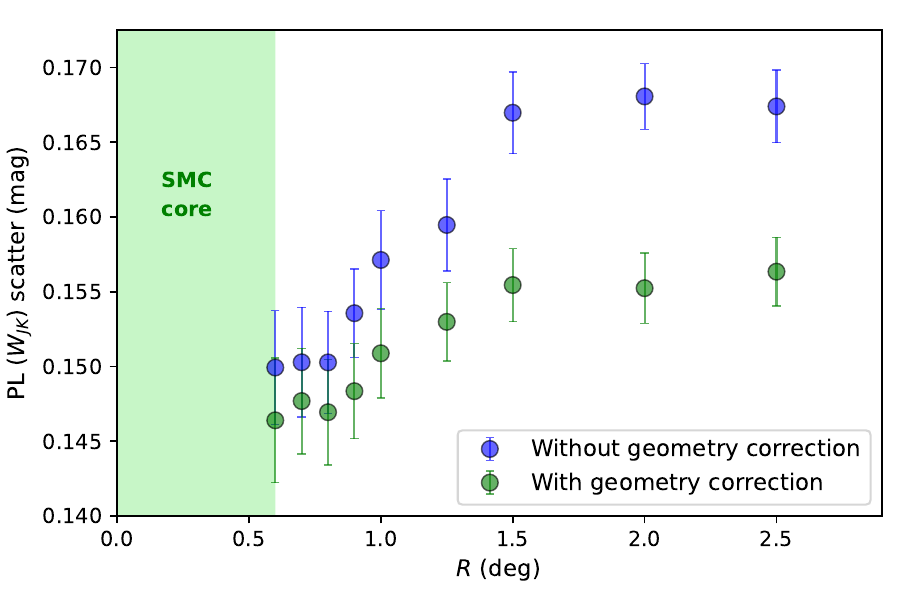} ~ 
\includegraphics[width=8.8cm]{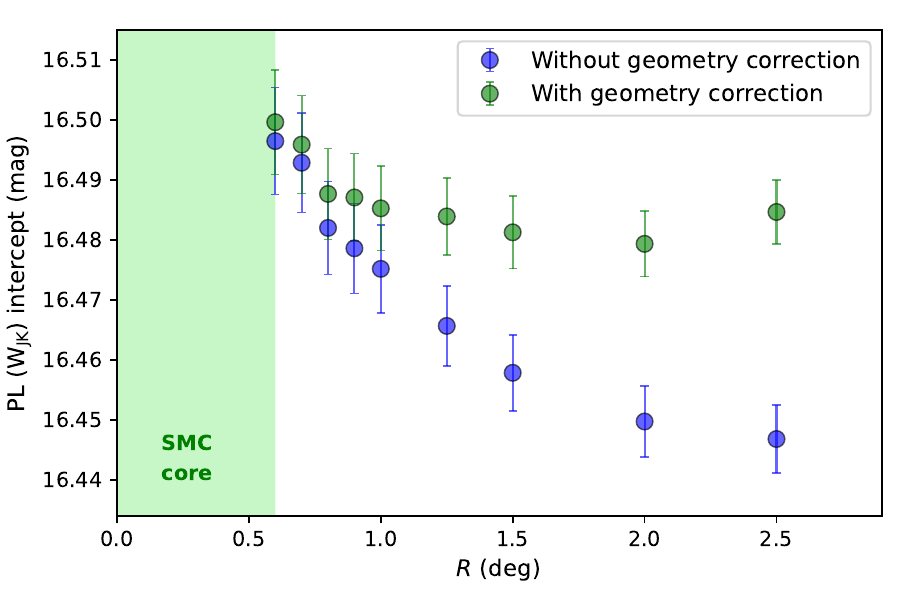} 
\caption{Metallicity effect ($\gamma$) (top left panel), PL scatter (top right panel) and PL intercept (bottom panel) in the $W_{JK}$ band as a function of the radius of the SMC region considered. The values are based on the dataset and method adopted by \citet{Wielgorski2017}. Blue points are the values found in the same conditions as \citet{Wielgorski2017} and green points represent the values obtained after correcting for the SMC geometry \citep{Graczyk2020}.}
\label{fig:scatter_W17}
\end{figure*}

If the SMC metallicity was to be revised to a more metal-poor value, this would not affect the main conclusion of the present work ($\gamma$ mostly independent of wavelength) and the updated $\gamma$ values would still be in excellent agreement with other findings from the literature.   \\

\subsection{Comparison with other empirical estimates of $\gamma$}
\label{sec:comp_lit}

The recognition of a $\gamma$ term is relatively recent owing to necessary improvements in data quality (i.e., parallaxes, reddenings, Cepheid photometry, metallicity and Cloud geometry) which have accrued in the last two decades. In this section, we aim at identifying sources of differences with some previous studies. We represent some of these previously published values of the metallicity effect in Fig. \ref{fig:gamma_time}.

In $W_{VI}$ our value of $\gamma = -0.201 \pm 0.071$ mag/dex agrees well with early estimates by \citet{Kennicutt1998}, \citet{Sakai2004}, \citet{Macri2006} and \citet{Scowcroft2009}. In all filters we find a stronger negative metallicity effect than \citet{Storm2011b} and \citet{Groenewegen2013} who both used Baade-Wesselink distances and report $\gamma$ values mostly consistent with zero. \citet{Gieren2018} also adopted a similar approach but measured a stronger effect (around $-0.27$ mag/dex), consistent with the present study although with larger error bars. \citet{Groenewegen2018} obtained a shallower metallicity effect in $V$, $K$ and $W_{VK}$ based on \textit{Gaia} DR2 parallaxes but still comparable with our findings within the error bars, similarly to \citet{Ripepi2020}. The results by \citet{Ripepi2021} based on \textit{Gaia} EDR3 parallaxes are close to our values in $K$, $W_{JK}$ and $W_{VK}$ in the sense that they show a strong effect, although their values are more negative by about 0.1 mag/dex. Recently, \citet{CruzReyes2022} compared a sample of MW open clusters with the LMC Leavitt law and obtained $\gamma$ values in $G$, $BP$, $RP$, $I$, $W_{VI}$, $W_G$ and $W_H$ that are in good agreement with our findings. Finally, in the $W_H$ Wesenheit index based on pure HST photometry of Milky Way and LMC Cepheids, we obtain a metallicity effect of $-0.280 \pm 0.078 \, \rm mag/dex$, in agreement with the value of $-0.217 \pm 0.046 \, \rm mag/dex$ derived by the SH0ES team from a broader range of data from the MW, LMC, SMC, NGC 4258 and gradients in SN Ia hosts \citep{Riess2022}.

\citet{Freedman2011} used spectroscopic [Fe/H] abundances of 22 individual LMC Cepheids whose metallicities were measured by \citet{Romaniello2008}, covering a range of about $0.6 \, \rm dex$. They derived a negative metallicity effect in the mid-IR, cancelling in the NIR and becoming positive in optical wavelengths. However, \citet{Romaniello2021} published new abundances for a larger sample of LMC Cepheids and report this time a very narrow distribution of metallicities ($\sigma = 0.1 \, \rm dex$ including systematics). They also note that the abundances provided in \citet{Romaniello2008} were significantly affected by a systematic error in the data reduction and analysis, and they confirm that the previous data are compatible with the same narrow spread observed for the new values. This shows that the LMC cannot be used to internally measure the metallicity effect and explains differences in the findings of \citet{Freedman2011} \citep[see][for further discussion]{Romaniello2021}.

\citet{Wielgorski2017} performed a purely differential calibration of the $\gamma$ term by comparing the Leavitt law across the full span of the LMC and SMC. Assuming the detached eclipsing binary distance by \citet{Pietrzynski2013} and \citet{Graczyk2014} respectively, they obtained a metallicity effect consistent with zero in the optical and NIR. Updating the mean LMC metallicity with the recent value by \citet{Romaniello2021} or/and replacing the DEB distances by the most precise ones by \citet{Pietrzynski2019} and \citet{Graczyk2020} does not yield significant differences with the \citet{Wielgorski2017} results. However, we find that the size of the region and its depth considered around the SMC center considerably impacts the value of the $\gamma$ term. The sensitivity of $\gamma$ to the size of the region adopted in the SMC is due to the elongated shape of this galaxy along the line of sight. Despite the geometry correction performed in Sect. \ref{sect:dist}, Cepheids at larger distance to the SMC center show a larger scatter, as shown in Fig. \ref{fig:scatter_W17}, likely due to the shortcoming of the planar model at greater radii (and farther from the region it was defined by the DEBs). Thus the distances to Cepheids in outer regions of the SMC may differ significantly from the mean SMC distance modeled from the DEBs, perhaps due to the structures of the Magellanic stream \citep{Nidever2008}.

We can reproduce the findings of \citet{Wielgorski2017} by neglecting the correction for the SMC geometry and when considering all SMC Cepheids ($R > 2 \, \rm deg$) for which we obtain the same low metallicity dependence as \citet{Wielgorski2017}. However the $\gamma$ term becomes more negative when we retain Cepheids in a smaller region and it reaches $-0.150 \, \rm mag/dex$ for $R < 0.6 \, \rm deg$. When the geometry of the SMC is included in apparent magnitudes, $\gamma$ is particularly stable between $R = 0.6 \, \rm deg$ and $R = 2 \, \rm deg$ with a value of $-0.150 \, \rm mag/dex$. This demonstrates that the very low metallicity effect found by \citet{Wielgorski2017} is likely due to unaccounted for differences in depth: limiting their analysis to a narrower region of the SMC and applying geometry corrections would yield a $\gamma$ value no longer consistent with zero. This issue was already mentioned by \citet{Gieren2018}. We note that the $\gamma$ values described in this section and represented in Fig. \ref{fig:scatter_W17} differ from our results listed in Table \ref{table:PLZ} since they are based on the data and method from \citet{Wielgorski2017} (i.e. LMC and SMC samples only).

\citet{Owens2022} (hereafter \citetalias{Owens2022}) compared Cepheids and geometric distances in the MW, LMC, and SMC and claim poor agreement, attributing this to an error in \textit{Gaia} EDR3 parallaxes and proposing a large, positive \textit{Gaia} parallax offset coupled with no Cepheid metallicity term (including for the commonly found one in $W_{VI}$), with the consequence of a shorter Cepheid distance scale and higher Hubble constant\footnote{The +18 $\mu$as \textit{Gaia} offset for bright objects proposed by \citetalias{Owens2022} conflicts with the $\sim$ -15 $\mu$as mean of the measurements external to \textit{Gaia} as summarized by \citet{Lindegren2021_plx_bias} and shown in Figure 1 by \citet{Li2022}. This offset also makes Cepheids in the OW22 fainter by 0.074 mag, the Clouds closer by that amount and {\it raises} the local Hubble constant and its present tension by $\sim$ 3.5\% without a metallicity term. In contrast, the metallicity term between the Clouds and MW presented here rather than a large, positive \textit{Gaia} offset provides the consistency between the \textit{Gaia} and DEB distances as they are the same size and direction. \\}. There are numerous, important differences in the data used by \citetalias{Owens2022} and ours: \citetalias{Owens2022} largely employs older and less consistently calibrated photometry and reddening estimates\footnote{For the MW NIR, \citetalias{Owens2022} photometry is from \citet{Welch1984}, \citet{Laney1992}, \citet{Barnes1997} whereas ours is from \citet{Monson2011} with the latter having twice as many Cepheids and better calibration. For MW reddenings, \citetalias{Owens2022} uses the \citet{Fernie1995} database, a literature compilation of photoelectric photometry from uncommon bandpasses with a mean era of the 1980's and the \citet{Cardelli1989} reddening law, products that have not benefitted from the modern wide-field studies from Pan-STARRS and SDSS like Bayestar \citet{Green2019}. For the LMC NIR data, \citetalias{Owens2022} uses 92 Cepheids from \citet{Persson2004} whereas this study augments that with $>$750 Cepheids from \citet{Macri2015}. For the SMC NIR data, \citetalias{Owens2022} uses data from \citet{Laney1986}, \citet{Welch1984} and \citet{Storm2004}, whereas we use a larger sample from \citet{Ripepi2016} and \citet{Kato2007}.} and a much smaller sample of MW Cepheids in the optical, 37 vs the $\sim$150 used here. We also use specific high-quality, space-based Cepheid photometry from HST (MW, LMC) and \textit{Gaia} (MW, LMC, SMC) not used in the \citetalias{Owens2022} study. It is beyond the scope of this study to analyze the impacts of the older and newer data samples but we would not be surprised if they produce systematic differences at the few hundreths of a magnitude level relevant to the $\sim 0.1$ mag effects of metallicity (e.g., between the LMC and SMC).  

However, we identify two specific differences in the measurements between the LMC and SMC which appear to impact the OW22 calculation and which are independent of \textit{Gaia} and its calibration. The difference in distance between the DEBs in the LMC and SMC is given by \citet{Graczyk2020} as 0.500 $\pm 0.017$ mag. The excess difference we find between the SMC core and the LMC (Table 4) averaged across all bands is 0.08 mag (SMC Cepheids are net fainter). For the 0.34 dex difference in Cloud metallicity, the metallicity term is then $\sim$ 0.24 mag per dex. \citetalias{Owens2022} give a best differential distance of 0.511 $\pm 0.056$ mag, for an excess of 0.01 mag, 0.07 mag smaller than found here and implying a negligible metallicity term in all bands (including the Wesenheit band, $W_{VI}$ which has generally been found to be $-0.2$ mag/dex, see Table \ref{table:gamma_lit}).  However, as Figure \ref{fig:scatter_W17} shows the depth of the SMC at large radii adds dispersion and reduces the apparent distance. The \citetalias{Owens2022} study uses SMC data from \citet{Scowcroft2016} with a radius from the core of up to 2 degrees. \citet{Scowcroft2016} also noted a large spatial dependence in Cepheid distance across the greater region of the SMC.  The combination of correcting for the geometry (of 0.03 mag, given but not applied in \citetalias{Owens2022}) and limiting to the core ($<$ 0.6 deg here) accounts for 0.06 mag of the 0.07 mag difference with our study and thus the difference between a moderate or negligible metallicity term in all bands. We think the known depth of the SMC and observed reduction in Cepheid PL scatter by correcting for the DEB-based geometry and limiting to the core where most of the DEBs are found provides a strong argument this yields a more accurate result. An additional 25\% increase in the metallicity term between the LMC and SMC comes from the decrease in the metallicity difference between the Clouds between \citet{Romaniello2008} used by \citetalias{Owens2022} and \citet{Romaniello2021} used here. \\

\begin{table*}[]
\small
\caption{Main improvements and updates between \citet{Breuval2021} and the present analysis.}
\centering
\begin{tabular}{ l  c  c }
\multicolumn{3}{c}{ }  \\
\hline
\hline
  & \citet{Breuval2021} & Present work  \\
\hline
Filters: & $V$, $I$, $J$, $H$, $K$ & $V$, $I$, $J$, $H$, $K$ \\
& & \textit{Gaia} $G$, $BP$, $RP$  \\
& & \textit{Spitzer} [3.6], [4.5]  \\
\hline
Wesenheit indices: & $W_{VI}$, $W_{JK}$ & $W_{VI}$, $W_{JK}$  \\
 & & $W_{VK}$, $W_{H}$ (HST), $W_{G}$ (Gaia)     \\
\hline
Reddening law: & $A_{\lambda}/A_V$ from \citet{Cardelli1989} & $A_{\lambda}/A_V$ from \citet{Fitzpatrick1999} \\
& & + uncertainties on $A_{\lambda}/A_V$ values \\
& & + uncertainties on $R_V$ ($3.1 \pm 0.1$) \\
\hline
\textit{Gaia} EDR3 parallax ZP: & \citet{Lindegren2021_plx_bias} & \citet{Lindegren2021_plx_bias}  \\
 & $\varpi = \varpi_0 - \rm ZP_{L21}$ & $\varpi = \varpi_0 - (\rm ZP_{L21} + 0.014 \, \mu as)$ \\
\hline
LMC metallicity: & [Fe/H] $= -0.34 \pm 0.06$ dex &  [Fe/H] $= -0.407 \pm 0.020$ dex \\
& \citep{Gieren2018} & \citep{Romaniello2021} \\
\hline
Reddening for MW Cepheids:  & \citetalias{Anderson2013}, \citetalias{Kovtyukh2008}, \citetalias{Laney2007}, \citetalias{Sziladi2007}, \citetalias{Acharova2012}, \citetalias{Fernie1995} & (a) Bayestar dust map \citep{Green2019}  \\
 &    & (b) Period-color relation \citep{Riess2022} \\
 &    & (c) SPIPS method \citep{Trahin2021} \\
\hline
 Reddening for LMC and & \citet{Gorski2020} & \citet{Skowron2021}  \\
SMC Cepheids:  & reddening maps & reddening maps  \\
\hline
$V$, $I$ photometry for & OGLE-IV \citep{Soszynski2015} & OGLE-IV \citep{Soszynski2015} \\
 LMC Cepheids & & + Shallow Survey \citep{Ulaczyk2013} \\
\hline
$G$, $BP$, $RP$ photometry & \textit{Gaia} DR2 light curves & \textit{Gaia} EDR3 light curves \\
 & \citep{Clementini2019} & \citep{Ripepi2022Gaia} \\ 
\hline
~ \\
\multicolumn{3}{l}{ \textbf{References}:  (A13): \citet{Anderson2013}, (K08): \citet{Kovtyukh2008}, (LC07): \citet{Laney2007},  }\\
\multicolumn{3}{l}{ (S07): \citet{Sziladi2007}, (A12): \citet{Acharova2012}, (F95): \citet{Fernie1995}.   }\\
~ \\
\end{tabular}
\label{table:B21_B22}
\end{table*}

\subsection{Comparison between empirical estimates and theoretical predictions}
\label{sec:comp_theory}

\begin{figure}[t!]
\centering
\includegraphics[width=7.6cm]{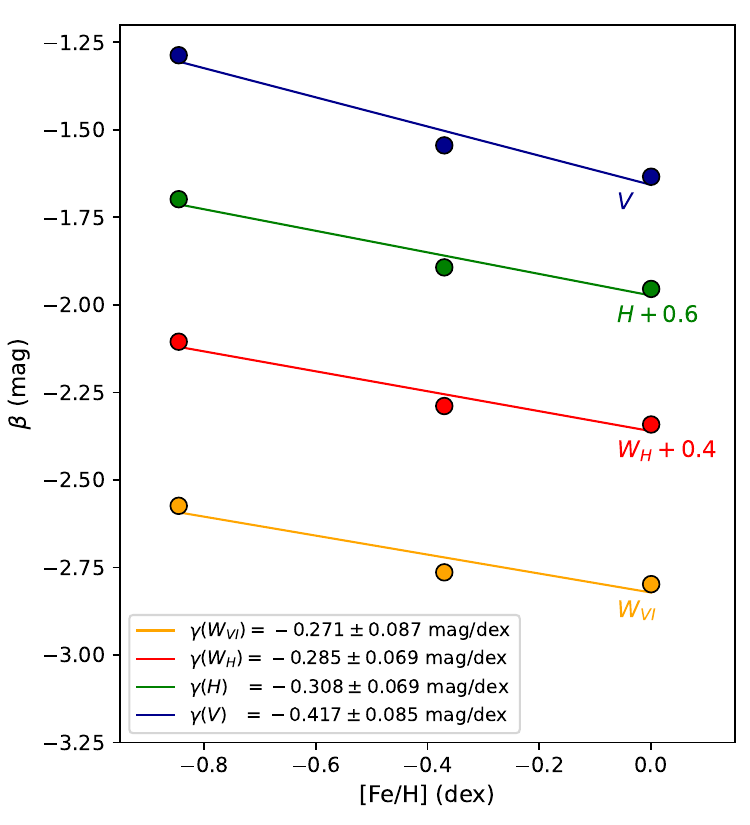} 
\caption{Metallicity effect predicted by \citet{Anderson2016} using the Geneva evolution models including rotation.  }
\label{fig:RIA2016}
\end{figure}

While empirical estimates of the metallicity term of the PL relation have become more precise due to better parallaxes \citep{GaiaEDR3_contents}, reddening estimates, Cepheid photometry, and knowledge of Cloud geometry, they may appear to conflict with earlier predictions from the theory based on non-linear convecting models \citep{Bono1999, Bono2008, Caputo2000, Marconi2005}. These studies suggested a positive sign for the $\gamma$ term, meaning that metal-rich Cepheids would be fainter. On the other hand, \citet{Anderson2016} recently performed a pulsation instability analysis of the linear Geneva stellar evolution models by \citet{Georgy2013} that include the effects of rotation. They predicted the PL relation in $V$, $H$, $W_{VI}$ and $W_H$ for three different metal abundances ($Z = 0.014$, $0.006$, and $0.002$, selected to match the MW, LMC and SMC Cepheids mean metallicity, respectively) and separately on the red and blue edge of the instability strip. We averaged the PL intercepts $\beta$ listed in Table 2 of \citet{Anderson2016} on both edges for the second and third crossing of the instability strip, and represented them with [Fe/H] on Fig. \ref{fig:RIA2016}. We find that the variation of these intercepts with [Fe/H] yield a negative metallicity effect of $\gamma \sim -0.27 \, \rm mag/dex$ to $-0.42 \, \rm mag/dex$ across the optical and NIR, consistent with our present results. Similarly, \citet{DeSomma2022} presented an extended set of nonlinear convective pulsation models for different metallicity values and also concluded with a negative metallicity term in different Wesenheit indices (see Table \ref{table:gamma_lit}), in agreement with \citet{Anderson2016}: these two theoretical studies best fit our observational data. Additionally, the \citet{Anderson2016} models reproduce particularly well the observed boundaries of the instability strip \citep{Groenewegen2020SEDs}. While additional theoretical studies are warranted, the agreement found here is quite promising. \\


\section{Conclusions and perspectives} 
\label{sec:conclusion}

The results by \citet{Breuval2021} suggested a possible dependence of $\gamma$ with wavelength, but were based on only five filters. To explore this question, we extended the wavelength range by including new mid-IR (\textit{Spitzer}) and optical (\textit{Gaia}) bands. We were able to take this analysis one step further by improving the technique and updating the data wherever it was possible. All the improvements included in this study are listed in Table \ref{table:B21_B22} and compared with the previous \citet{Breuval2021} paper. We note that the uncertainties on the $\gamma$ terms presented in this paper are not significantly smaller compared with previous analysis, despite the several improvements included: this is because we now include the uncertainties on the reddening law and on the $R_V$ values, which were not considered previously.

We report values of the metallicity effect on the Cepheid PL relation in 10 filters from $0.5 \, \rm \mu m$ to $4.5 \, \rm \mu m$ and in 5 Wesenheit indices, including the HST based Wesenheit index $W_H$ used for the SH0ES distance ladder \citep{Riess2022}. We obtain a negative $\gamma$ term in all bands, meaning that metal-rich Cepheids are brighter than metal-poor ones, in agreement with all recent empirical studies. We find a globally uniform value of $\gamma$ of about $-0.28 \, \rm mag/dex$ from optical to mid-IR filters, showing that the main influence of metallicity on Cepheids is in their brightness rather than color. 

While our results are largely consistent with recent measurements, we track differences in two studies, \citep{Wielgorski2017,Owens2022} that employ the SMC to a depth effect.  Correcting for the geometry and limiting the radius to the SMC core is shown to narrow the distance range resulting in a sample insured to be at the same distance as the DEBs that also produces a metallicity term on the same trend line as seen between the MW and LMC.

Comparing Cepheids over a sufficiently large metallicity range still requires to combine different samples of Cepheids located in several galaxies, having different distances, photometric systems, dust distribution and properties (e.g. reddening law), which implies large systematic uncertainties. 
In the near future, it should be possible to reduce the impact of these systematics and to increase the precision of the $\gamma$ term thanks to the 4th \textit{Gaia} data release. Ideally, these new distance measurements will have to be combined with consistent metallicity estimates of all Milky Way Cepheids obtained in a single system, spanning a wide range of abundances. In this sense, improvements are also expected from the use of recently published \citep{Ripepi2021, Romaniello2021, DaSilva2022} and upcoming abundance catalogs for MW, LMC and SMC Cepheids, which should again help to calibrate the metallicity effect with a better accuracy. \\

\newpage
\section*{Acknowledgements}

We thank the referee for their constructive report which helped to improve the present paper. L.B. is grateful to C. A. L. Bailer Jones, F. Ar\'enou, B. Trahin, L. M. Macri, S. Casertano and A. M\'erand for inspiring discussions which helped to improve the present work. We thank L. M. Macri for providing the photometric transformations to the 2MASS system. We are grateful to V. Ripepi for providing the full list of reclassified Cepheids with \textit{Gaia} DR3. The research leading to these results  has received funding from the European Research Council (ERC) under the European Union's Horizon 2020 research and innovation program (projects CepBin, grant agreement 695099, and UniverScale, grant agreement 951549). This work has made use of data from the European Space Agency (ESA) mission {\it Gaia} (\url{https://www.cosmos.esa.int/gaia}), processed by the {\it Gaia} Data Processing and Analysis Consortium (DPAC, \url{https://www.cosmos.esa.int/web/gaia/dpac/consortium}). Funding for the DPAC has been provided by national institutions, in particular the institutions participating in the {\it Gaia} Multilateral Agreement. The results presented in this paper benefited from discussions with the International Space Science Institute (ISSI) team led by G. Clementini\footnote{\href{https://www.issibern.ch/teams/shot/}{https://www.issibern.ch/teams/shot/}}. This research has made use of Astropy, a community-developed core Python package for Astronomy \citep{2018AJ....156..123A}. We used the SIMBAD and VIZIER databases and catalog access tool at the CDS, Strasbourg (France), and NASA's Astrophysics Data System Bibliographic Services. This research has made use of the SVO Filter Profile Service\footnote{\href{http://svo2.cab.inta- csic.es/theory/fps/}{http://svo2.cab.inta- csic.es/theory/fps/}}. Some of the data presented in this paper were obtained from the Mikulski Archive for Space Telescopes (MAST) at the Space Telescope Science Institute. The specific observations analyzed can be accessed via \dataset[10.3847/1538-4357/ab1422]{https://doi.org/10.3847/1538-4357/ab1422} and \dataset[10.3847/2041-8213/abdbaf]{https://doi.org/10.3847/2041-8213/abdbaf}.\\

\newpage
 \appendix

 \begin{figure}[h!]
\centering
\includegraphics[width=15.0cm]{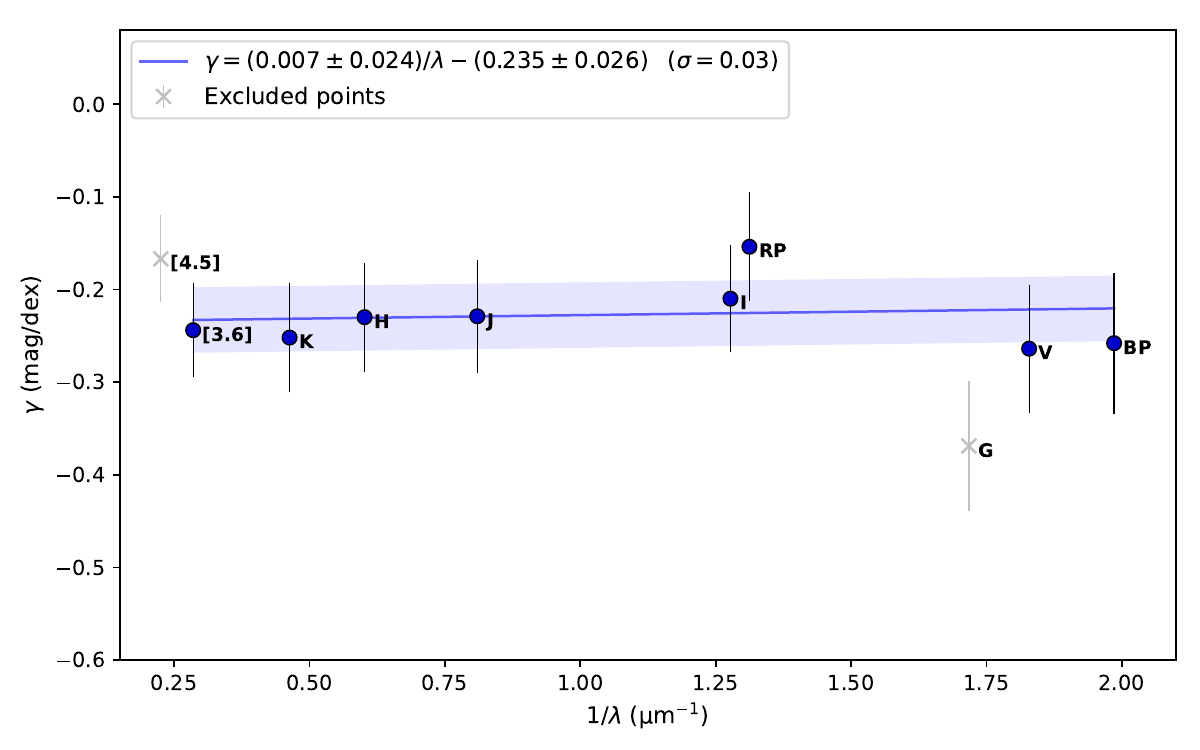} 
\caption{Metallicity effect ($\gamma$) as a function of the inverse of wavelength ($1/\lambda$) in the hypothesis of a more metal-poor SMC sample of $-0.90 \pm 0.05$ dex (see discussion in Sect. \ref{sec:SMC_systematics}).}
\label{fig:gamma_lambda_metalpoor}
\end{figure}

\bibliography{Breuval_2022}{}
\bibliographystyle{aasjournal}


\end{document}